\documentclass[epj,nopacs]{svjour}
\usepackage[T1]{fontenc}
\usepackage{wrapfig}
\usepackage[caption=false, singlelinecheck=false, justification=raggedright]{subfig}
\usepackage{float}
\usepackage{array}
\usepackage{booktabs}
\usepackage{feynmp}
\usepackage{ulem}
\usepackage{soul,color}
\usepackage{amsmath}
\usepackage{amssymb}
\usepackage{amsfonts}
\usepackage{mathrsfs}
\usepackage{wasysym}
\usepackage{upgreek}
\usepackage{multirow}
\usepackage[table]{xcolor}
\usepackage{nicefrac}
\usepackage{extarrows}
\usepackage{txfonts}
\usepackage[scaled=0.92]{helvet}
\usepackage{paralist}
\usepackage{slashed}
\usepackage{tikz}
\usepackage{enumitem}
\usepackage{cancel}
\usepackage{cite}%
\newcommand{\modulus}[1]{\ensuremath{\left| #1 \right|}}
\renewcommand{\Re}{\textrm{Re}}
\renewcommand{\Im}{\textrm{Im}}
\newcommand{\SM}{\textrm{SM}}

\newcommand{\be}{\begin{eqnarray*}}
\newcommand{\ee}{\end{eqnarray*}}
\newcommand{\bea}{\begin{eqnarray}}
\newcommand{\eea}{\end{eqnarray}}

\journalname{Eur. Phys. J. C}
\def\makeheadbox{\relax}
\allowdisplaybreaks

\begin{document}
\title{Probing the non-standard neutrino interactions using quantum
  statistics}

\author{C. S. Kim\inst{1,2}\thanks{Email at: cskim@yonsei.ac.kr},
  Janusz~Rosiek\inst{3}\thanks{Email at: Janusz.Rosiek@fuw.edu.pl},
  \and Dibyakrupa Sahoo\inst{3}\thanks{Email at:
    Dibyakrupa.Sahoo@fuw.edu.pl}}

\authorrunning{Kim, Rosiek and Sahoo}

\institute{Department of Physics and IPAP, Yonsei University, Seoul 03722, Korea 
\and Institute of High Energy Physics, Dongshin University, Naju 58245, Korea
\and Faculty of Physics, University of Warsaw, Pasteura 5, 02-093 Warsaw, Poland}%

\date{\today}
\abstract{Using the well established principles of Lorentz invariance,
  CP and CPT symmetry, and quantum statistics we do a
  model-independent study of effects of possible non-standard
  couplings of (Dirac and Majorana) neutrinos.  The study is sensitive
  to the different quantum statistical properties of the Dirac and
  Majorana neutrinos which, contrary to neutrino-mediated processes of
  lepton number violation, could lead to observable effects
  \textit{not} suppressed by the small ratios of neutrino and heavier
  particle masses.  For processes with a neutrino-antineutrino pair of
  the same flavor in the final state, we formulate the ``Dirac
  Majorana confusion theorem (DMCT)'' showing why it is normally very
  difficult to observe the different behaviour of both kinds of
  neutrinos in experiments if they have only the standard model
  (SM)-like left-handed vector couplings to gauge bosons.  We discuss
  deviations from the confusion theorem in the presence of
  non-standard neutrino interactions, allowing to discover or
  constrain such novel couplings.  We illustrate the general results
  with two chosen examples of neutral current processes, $Z \to \nu \,
  \overline{\nu}$ and $\mathcal{P}_i \to \mathcal{P}_f \, \nu \,
  \overline{\nu}$ (with $\mathcal{P}_{i,f}$ denoting pseudoscalar
  mesons, such as $B,K,\pi$). Our analysis shows that using 3-body
  decays the presence of non-standard interactions can not only be
  constrained but one can also distinguish between Dirac and Majorana
  neutrino possibilities.  }

\maketitle

\section{Introduction}\label{sec:introduction}

Observations of neutrino oscillation \cite{Fukuda:1998mi,Ahmad:2002jz}
have established the fact that neutrinos have non-zero but tiny masses
and the flavor neutrinos ($\nu_\ell$ with $\ell=e, \mu, \tau$), which
only participate in the weak interactions, are linear combinations of
mass eigenstates ($\nu_i$ with $i=1, 2, 3$ having masses $m_i$),
\begin{equation}
\left\lvert \nu_\ell \right\rangle = \sum_i U_{\ell i}^* \left\lvert
\nu_i \right\rangle,
\end{equation}
where $U$ denotes the unitary $3\times 3$ lepton mixing matrix, also
called the PMNS matrix \cite{Maki:1962mu, Pontecorvo:1967fh}.  At
present \cite{ParticleDataGroup:2020ssz}, we know neither the values
of the individual masses of the neutrinos nor the mechanism that gives
rise to these tiny masses (see Refs.~\cite{deGouvea:2016qpx,
  Cai:2017jrq} for recent reviews of various neutrino mass models). In
the Standard Model of particle physics (SM) neutrinos are massless and
get produced via weak interaction such that all neutrinos are
left-handed and all antineutrinos are right-handed.  However, as the
neutrinos are \textit{not} massless, handedness is not a good quantum
number for neutrinos, i.e.\ in a frame moving faster than the neutrino
(or antineutrino) the handedness will be reversed.  Nevertheless,
direct production of right-handed neutrinos and left-handed
antineutrinos has not yet been observed in any experiment, consistent
with the SM allowed $V-A$ interactions.  Therefore, it is currently
unknown whether the right-handed neutrinos have the same mass as the
left-handed neutrinos or not.  Most certainly, detection of the
right-handed neutrinos and the left-handed antineutrinos would require
invoking some sort of new physics (NP) which we currently do not know.
To complicate the matter further, neutrinos being devoid of any
electric or colour charge are the only known fermions which can
\textit{possibly} be their own antiparticles, i.e.\ they can be
Majorana fermions \cite{Majorana:1937vz, Racah:1937qq, Furry:1938zz}
unlike all the others which are Dirac fermions.\footnote{In contrast,
massless neutrinos are described as Weyl fermions. The reduction of
neutrino degrees of freedom from 4 to 2 for massless neutrinos is a
discrete jump. Thus massless neutrinos are completely different from
massive neutrinos even with an extremely tiny mass.  In this work the
mass of neutrinos is always taken to be non-zero.} In most of the
interesting mass models for neutrinos, all of which involve different
NP setups, the Majorana nature is assumed.  Like with many other
aspects of neutrinos, this is yet to be decisively decided by
experiments.

Various methodologies that have been proposed to probe the Majorana
nature of neutrinos can be broadly grouped into the following two
categories, depending on the type of process being considered.
\begin{enumerate}
\item \textit{Propagator probes} or \textit{neutrino mediated probes}:
Well known examples of such processes are neutrinoless double-beta
decay ($0\nu\beta\beta$) \cite{Avignone:2007fu}, neutrinoless
double-electron capture \cite{Blaum:2020ogl}, neutrinoless muon
($\mu^-$) to positron ($e^+$) conversion \cite{Pontecorvo:1967fh,
Lee:2021hnx}, $\nu$ exchange force ($\nu$ Casimir force)
\cite{Grifols:1996fk, Ferrer:1998ju, Segarra:2020rah,
Costantino:2020bei, Ghosh:2019dmi, Bolton:2020xsm, Xu:2021daf,
Ghosh:2022nzo} or lepton number violating massive sterile neutrino
exchange \cite{Cvetic:2010rw}, which involve one or more neutrino(s)
and/or antineutrino(s) as propagator(s). The lepton number violating
processes involve helicity flip and are, therefore, always
proportional to the unknown mass of the propagating Majorana neutrino.
In addition, all the nuclear processes explored as propagator probes
to infer the nature of sub-eV active neutrinos suffer from large
theoretical uncertainty in the nuclear matrix elements.  Although
$0\nu\beta\beta$ searches have thus far not yielded any signal
\cite{CUORE:2021mvw}, the experimental searches for it are going on.
Nevertheless, the associated issues mentioned above underline the
importance of alternative methods.%

\item \textit{Initial / final state probes}: These involve one (or
more) neutrino(s) and/or antineutrino(s) in the initial state or the
final state.  Some previously explored processes under this category
are the neutrino-electron elastic scattering \cite{Kayser:1981nw,
Rosen:1982pj, Dass:1984qc, Rodejohann:2017vup}, neutrino-nucleon
scattering \cite{Kayser:1981nw}, neutrino-nuclei scattering
\cite{Millar:2018hkv}, 2-body processes $\gamma^* \to
\nu\,\overline{\nu}$ \cite{Kayser:1982br} that try to explore the
electromagnetic properties of neutrinos \cite{Schechter:1981hw}, $Z
\to \nu \, \overline{\nu}$ \cite{Shrock:1982jh}, $e^+ \, e^- \to \nu
\, \overline{\nu}$ \cite{Ma:1989jpa}, 3-body processes $K^+ \to
\pi^+\,\nu\,\overline{\nu}$ \cite{Nieves:1985ir}, $e^+ \, e^- \to
\gamma\,\nu\,\overline{\nu}$ \cite{Chhabra:1992be}, $e\,\gamma \to e
\, \nu \, \overline{\nu}$ \cite{Berryman:2018qxn}, radiative emission
of neutrino pair \cite{Yoshimura:2013wva}, and 4-body $B$ meson decay
$B^0 \to \mu^+ \, \mu^- \, \nu_\mu \, \overline{\nu}_\mu$
\cite{Kim:2021dyj}.  In contrast with the propagator probes which
essentially directly test the Majorana mass term, the initial/final
state probes try to examine the quantum mechanical identicalness of
Majorana neutrino and antineutrino via quantum statistics.

Unlike the neutrino mass suppressed processes for propagator probes,
the initial/final state probes involve SM allowed processes and
therefore can be produced relatively easily in experiments. The
challenge in probing the Dirac or Majorana nature of neutrinos using
such probes is to infer the momenta or energies of the missing
neutrinos which is always experimentally challenging.
\end{enumerate}

The quantum statistics of Majorana neutrino and antineutrino stems
from their quantum mechanical identical nature, and it is an intrinsic
property of the Majorana neutrino just like its spin and charge.  For
a final state containing a pair of Majorana neutrino and antineutrino
of the same flavor, the Fermi-Dirac statistics asserts that the
corresponding transition amplitude is always anti-symmetrized with
respect to their exchange.  This anti-symmetrization does not depend
on the size of the mass of neutrino ($m_\nu$).  Therefore, there is no
reason \textit{a priori} to think that the difference between Dirac
and Majorana neutrinos via such initial/final state probes would
necessarily be dependent on $m_\nu$, a result that apparently (but not
necessarily) contradicts the ``\textit{practical} Dirac Majorana
confusion theorem'' (DMCT) \cite{Kayser:1981nw, Kayser:1982br}.  A
brief overview of the literature, in this context, comparing two-body
decays, three-body decays as well as four-body decays is given in
Ref.~\cite{Kim:2021dyj}.

The various initial/final state probes can, in general, be related to
one another via crossing symmetry~\cite{Rosen:1982pj}.  For ease of
discussion and without loosing any generality we concentrate in this
paper on providing a generalised framework for final state probes,
i.e.\ using processes in which a pair of neutrino and antineutrino
would appear in the final state.  Furthermore, we relax the helicity
considerations, i.e.\ we do not confine ourselves to left-handed
neutrinos and right-handed antineutrinos alone, since the involved
interactions will take care of the helicities by default.  Unlike
considering only SM allowed interactions as done in
Ref.~\cite{Kim:2021dyj} here we consider NP effects in a
model-independent manner and point out their effects as well. 

Before providing the layout of our paper, we would like to emphasise
that the active neutrinos and antineutrinos in all the final states
considered in this paper are invisible. In many NP models it is
possible to have other invisible particles such as the heavy sterile
neutrinos, dark matter particles or some long-lived light particles
(e.g. axion-like) that decays outside the detector volume. Therefore,
observation of decays into invisible states with rates modified
comparing to the SM predictions usually is not sufficient by itself to
determine the source of the modification, being it different types of
neutrinos or entirely novel particles. The analysis we present in this
paper is limited by the assumptions that neutrinos are only invisible
final state but it is still useful for several reasons.
\begin{itemize}
\item If new invisible particles are massive (even relatively light
but with masses significantly exceeding neutrino masses), they could
have noticeable effects such as significant reduction in phase space,
or presence of some resonance peaks corresponding to on-shell
production of the long-lived particles. Corresponding modifications of
visible final state particle spectra could help in distinguishing such
cases from the neutrino production. Such possibilities are well
explored in the literature, see e.g. \cite{Goudzovski:2022vbt} for a
recent detailed review of invisible $K$ meson decays in various SM
extensions such as the 2-body decay $K \to \pi \, X_{\textrm{inv}}$
where the invisible particle $X_{\textrm{inv}}$ could be a Higgs mixed
scalar, dark photon, or axion-like particle.%
\item Current experiments are in general in agreement with the SM
predictions, therefore the bounds on non-standard neutrino couplings
which we discuss in the paper come from comparing their effects with
the size of experimental errors of relevant observables.  If there are
more invisible states in a given NP model, even very light ones,
barring fine tuning they compete with neutrino interactions in
saturating the errors, tightening the bounds. Thus the limits on NP
neutrino couplings which we discuss can be considered as (again
barring fine tuning and potential cancellations) being on the
conservative side. %
\item Finally, in many SM extensions neutrinos remain the only
ultra-light and weakly interacting (thus ``invisible'') particles. In
such cases methodology proposed in the paper can be directly used to
constrain NP neutrino couplings for a chosen BSM model.
\end{itemize}

Our paper is organised as follows.  Following Introduction, in
section~\ref{sec:formalism} we present a model-independent general
formalism for processes involving a neutrino and antineutrino pair of
the same flavor in the final state and we provide a generalised
statement of DMCT in context of these processes.  In
section~\ref{sec:np} we discuss how the effects of quantum statistics
and symmetry properties of transition amplitudes can be utilised to
probe NP interactions in neutrino sector and potentially lead to
substantial difference between Dirac and Majorana neutrinos.  In the
following section~\ref{sec:examples} we illustrate the general results
with two suitably chosen two-body and three-body neutral current
decays, showing how the bounds on the NP couplings depend on the
assumed neutrino nature.  The possibility of distinguishing the Dirac
and Majorana neutrinos by measuring the visible particles energy
spectrum in the three-body decays is discussed in
section~\ref{sec:dmdiff}.  Finally we conclude in
section~\ref{sec:conclusion} highlighting the various salient aspects
of our study.

\section{Model-independent formalism for processes containing $\boldsymbol{\nu\, \overline{\nu}}$ in the final state}
\label{sec:formalism}

\subsection{Choice of process}

Let us consider a general process with a neutrino and an antineutrino
of the same flavor in the final state, say $$X (p_X) \to Y(p_Y) \, \nu
(p_1) \, \overline{\nu}(p_2),$$ where $X, Y$ can be single or
multi-particle states, $Y$ can also be null, the contents of $X$ and
$Y$ (if it exists) can only be any visible particle and the 4-momenta
$p_X, p_Y$ are assumed to be well measured so that one can
unambiguously infer the total missing 4-momentum of $\nu\,
\overline{\nu}$, $p_\textrm{miss} = p_1 + p_2$.  So the 4-momentum of
$X$ must either be fixed by design of the experiment (e.g.\ $X$ might
be a particle produced at rest in the laboratory or be the constituent
of a collimated beam of known energy or it could consist of two
colliding particles of known 4-momenta) or the 4-momentum of $X$ be
inferred from the fully-tagged partner particle with which it is
pair-produced.  The final state $Y$ should not contain any additional
neutrinos or antineutrinos.  The process could be a decay or
scattering depending on whether $X$ is a single particle state or two
particle state.  Some actual processes that satisfy such criteria are
$e^+\, e^- \to \nu \, \overline{\nu}$, $Z \to \nu \, \overline{\nu}$,
$e^+\, e^- \to \gamma \, \nu \, \overline{\nu}$, $K \to \pi \, \nu \,
\overline{\nu}$, $B \to K \, \nu \, \overline{\nu}$, $B^0 \to \mu^+ \,
\mu^- \, \nu_\mu \, \overline{\nu}_\mu$, $J/\psi \to \mu^+ \, \mu^- \,
\nu_\mu \, \overline{\nu}_\mu$, $H \to \tau^+ \, \tau^- \, \nu_\tau \,
\overline{\nu}_\tau$ etc.

\begin{figure}[!h]
\centering %
\subfloat[Dirac neutrinos]{\includegraphics[width=0.7\linewidth,
keepaspectratio]{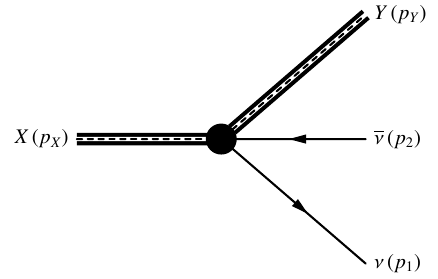}} %
\\ %
\subfloat[Majorana neutrinos (with exchange
diagram)]{\includegraphics[width=0.7\linewidth,
keepaspectratio]{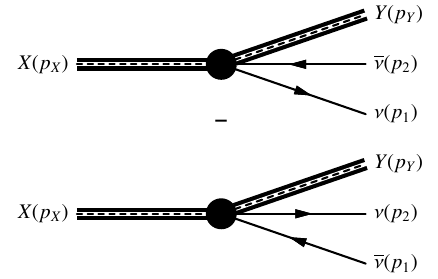}} %
\caption{A cartoon (pseudo-Feynman diagrams) for the processes $X \to
  Y \, \nu \, \overline{\nu}$, with both Dirac and Majorana neutrino
  possibilities.  The blob in these diagrams represents the effective
  vertex and it includes both SM and NP contributions.  Although
  Majorana antineutrino is indistinguishable from Majorana neutrino,
  we keep using the notation of $\overline{\nu}$ for antineutrino and
  $\nu$ for neutrino simply as a book-keeping device.} %
\label{fig:X2Ynn}
\end{figure}

Schematically the process $X \to Y \, \nu \, \overline{\nu}$ is drawn
in Fig.~\ref{fig:X2Ynn}.  Since Majorana neutrino and antineutrino are
quantum mechanically identical, there is an additional diagram with
exchanged 4-momenta $p_1 \leftrightarrow p_2$ in the Majorana case as
shown in Fig.~\ref{fig:X2Ynn}. The effect of this exchange diagram in
Majorana case can, in many instances but not always, be absorbed into
a single diagram like the one for the Dirac case but with a different
effective vertex.  In section~\ref{sec:examples} we shall encounter a
few examples where the effective vertex for Majorana case is different
from Dirac case.  An example process where we can not absorb the
direct and exchange diagram contributions to the effective vertex
factor is $B^0 \to \mu^+ \, \mu^- \, \nu_\mu \, \overline{\nu}_\mu$ as
discussed in Ref.~\cite{Kim:2021dyj}.

Additionally, it should be noted that in this work we assume that in
any of the processes under our consideration, the effect of
measurements should not destroy the identical nature of Majorana
neutrino and antineutrino.  This is akin to putting the constrain that
in a double-slit experiment meant to observe the interference of
light, no measurement should identify through which slit the photon
has passed.

\subsection{Origin of observable difference between Dirac and Majorana
  neutrinos}

The transition amplitude is, in general, dependent on all the
4-momenta of participating particles.  For brevity of expression and
without loss of generality, we denote the transition amplitude by only
mentioning the $p_1$, $p_2$ dependence.  For Dirac neutrinos, the
transition amplitude can be written as, %
\begin{equation}
\label{eq:amp-D}
\mathscr{M}^D = \mathscr{M}(p_1, p_2),
\end{equation}
while for Majorana case the amplitude is anti-symmetrized with respect
to the exchange of the Majorana neutrino and antineutrino which are
quantum mechanically identical fermions,
\begin{equation}
\label{eq:amp-M}
\mathscr{M}^M = \dfrac{1}{\sqrt{2}} \Big(\underbrace{\mathscr{M}(p_1,
  p_2)}_\textrm{Direct amplitude} - \underbrace{\mathscr{M}(p_2,
  p_1)}_\textrm{Exchange amplitude} \Big),
\end{equation}
where $1/\sqrt{2}$ takes care of the symmetry factor.  Note that the
amplitudes of Eqs.~\eqref{eq:amp-D} and \eqref{eq:amp-M} do not
necessarily assume SM interactions, they can involve NP effects as
well, and hence they include the most general structures of the
amplitude that are allowed by Lorentz invariance.  A general analysis
that specifically exploits the symmetry properties of the amplitude in
context of generic NP possibilities is given in Sec.~\ref{sec:np}.

The difference between Dirac and Majorana cases that can possibly be
probed is obtained after squaring the amplitudes and taking their
difference, which is given by,
\begin{align}
\left|\mathscr{M}^D\right|^2 -\left|\mathscr{M}^M\right|^2 &=
\frac{1}{2} \Bigg(\underbrace{\left|\mathscr{M}(p_1,
  p_2)\right|^2}_\textrm{Direct term} -
\underbrace{\left|\mathscr{M}(p_2, p_1)\right|^2}_\textrm{Exchange
  term}\Bigg) \nonumber\\%
&\quad + \underbrace{\textrm{Re}\left(\mathscr{M}(p_1, p_2)^*
  \, \mathscr{M}(p_2, p_1)\right)}_\textrm{Interference term}. \label{eq:general-D-M}
\end{align}
From Eq.~\eqref{eq:general-D-M} it is easy to conclude that there are
essentially two sources of difference between Dirac and Majorana
cases,
\begin{enumerate}
\item unequal contributions from ``Direct term'' and ``Exchange
  term'', i.e.\
\begin{equation}
\label{eq:direct-exchange-nonequal}
\underbrace{\left|\mathscr{M}(p_1, p_2)\right|^2}_\textrm{Direct term}
\neq \underbrace{\left|\mathscr{M}(p_2,
  p_1)\right|^2}_\textrm{Exchange term},
\end{equation}
\item non-zero contribution from the ``Interference term'', i.e.\
\begin{equation}
\label{eq:nonzero-interference}
\underbrace{\textrm{Re}\left(\mathscr{M}(p_1, p_2)^* \,
  \mathscr{M}(p_2, p_1)\right)}_\textrm{Interference term} \neq 0.
\end{equation}
\end{enumerate}
However, note that in the case when \textit{no individual information
  about $\nu\, \overline{\nu}$ are either known or deducible}, the
only difference between Dirac and Majorana cases that can be
experimentally accessed is obtained after full integration over $p_1$
and $p_2$ which gives,
\begin{align}
& \iint \left( \left|\mathscr{M}^D\right|^2
-\left|\mathscr{M}^M\right|^2 \right) \mathrm{d}^4 p_1 \, \mathrm{d}^4
p_2 \nonumber\\%
&= \iint \underbrace{\textrm{Re}\left(\mathscr{M}(p_1, p_2)^* \,
  \mathscr{M}(p_2, p_1)\right)}_\textrm{Interference term}
\mathrm{d}^4 p_1 \, \mathrm{d}^4 p_2,
\label{eq:integrated-D-M}
\end{align}
where we have used the fact that although, in general, the ``Direct''
and ``Exchange'' terms may differ, when we fully integrate over the
4-momenta of neutrino and antineutrino we do get %
\begin{equation}
\label{eq:DirectExchangeSame}
\iint \underbrace{\left|\mathscr{M}(p_1,
  p_2)\right|^2}_{\textrm{Direct term}} \mathrm{d}^4p_1 \,
\mathrm{d}^4 p_2 = \iint \underbrace{\left|\mathscr{M}(p_2,
  p_1)\right|^2}_{\textrm{Exchange term}} \mathrm{d}^4p_1 \,
\mathrm{d}^4 p_2,
\end{equation}
as $p_1$ and $p_2$ act as dummy variables in this case.

\subsection{``Dirac-Majorana Confusion Theorem''
  for the neutrino interactions in the SM }
\label{sec:dmct}

In most of the experimental scenarios, especially true for processes
of the form $X \to Y \, \nu \, \overline{\nu}$, information about
individual neutrino momenta is not available.  In such a case the
difference between the Dirac and the Majorana neutrinos is given only
by the integrated interference term in Eq.~\eqref{eq:integrated-D-M}.
If we further consider only SM interactions without any NP
contributions, then the $V-A$ interaction of SM would always produce
left-handed neutrino and right-handed antineutrino, even if the mass
of the neutrino and antineutrino is considered to be non-zero.  In
such a case the evaluation of the squared Feynman diagram for the
``Interference term'' would, in general, necessarily involve two
helicity flips which would make it proportional to $m_\nu^2$, i.e.\
\begin{equation*}
\underbrace{\textrm{Re}\left(\mathscr{M}^\SM(p_1, p_2)^* \,
  \mathscr{M}^\SM(p_2, p_1)\right)}_\textrm{Interference term in the
  SM} \propto m_\nu^2.
\end{equation*} Thus,
\textit{if only SM interactions are considered and one fully
  integrates over the neutrino and antineutrino 4-momenta, the
  difference between Dirac and Majorana cases would be proportional to
  $m_\nu^2$}.  This can be considered as a generalised statement of
the ``practical Dirac Majorana confusion theorem'' (DMCT).

The above statement still holds taking into account that in many cases
experiments can measure the total 4-momentum of the neutrino and
antineutrino pair in the final state, $p_1+p_2$, equal to the
difference between initial and final 4-momenta of the visible
particles.  The knowledge of $p_1+p_2$ only (not $p_1$ and $p_2$
individually), gives access to that part of the full phase space for
the neutrino pair which is obviously symmetric under the exchange $p_1
\leftrightarrow p_2$.  Thus argument about symmetric integration and
corresponding cancellation between ``Direct'' and ``Exchange'' terms
in Eq.~(\ref{eq:DirectExchangeSame}) still holds, even if the phase
space integral is not done over the full range of $p_1,p_2$.  In such
a scenario, all the cross sections or decay rate distributions
expressed in terms of parameters related to the 4-momenta of visible
particles will always be identical if pure SM interactions are assumed
(up to the neutrino mass suppressed effects).  However, as we will
show in the following sections, the distributions of visible particles
may differ if one also allows non-standard neutrino interactions.

The remaining possibility of experimental observation of the
difference between the Dirac and Majorana neutrinos in the SM is to
reconstruct or at least constrain the neutrino 4-momenta or some of
their components such as the energies, so that full integration over
the neutrino and antineutrino 4-momenta is not necessary and one can
directly use Eq.~(\ref{eq:general-D-M}), without worrying about the
cancellation in Eq.~(\ref{eq:DirectExchangeSame}).  This, in general,
is difficult as neutrinos are invisible in detectors and their
4-momenta can be only inferred from the kinematics of visible
particles.  Still, it may be possible for some particular
configuration(s) of them, however occurring only for a small fraction
of all kinematically allowed final states.

Relevant example of such ``special'' kinematic scenario has been
proposed in Ref.~\cite{Kim:2021dyj}, where authors consider the 4-body
decay of the form $X \to y_1\, y_2\, \nu\, \overline{\nu}$ with $y_1$
and $y_2$ flying away back-to-back with equal magnitudes of 3-momenta
in the rest frame of $X$.  In such case, the $\nu$ and
$\overline{\nu}$ also fly away back-to-back with equal energies
$E_\nu= (m_X - E_{y_1} - E_{y_2})/2$, where $E_{y_1}$, $E_{y_2}$
denote the energies of $y_1$ and $y_2$ respectively and $m_X$ is the
mass of parent particle $X$.  This is discussed in detail in
Ref.~\cite{Kim:2021dyj} taking the example of $B^0 \to \mu^- \, \mu^+
\, \nu_\mu \, \overline{\nu}_\mu$.  As the authors point out, other
meson decays such as those of neutral $K$, or $D$, or $J/\psi$, or
$\Upsilon(n\textrm{S})$ decays as well as Higgs decays might also be
useful for such studies.  The difference between Dirac and Majorana
cases does not necessarily depend on the size of the neutrino mass,
and hence might be feasible to discover in future experiments.

\section{New Physics scenarios and practical DMCT}
\label{sec:np}

There is no reason \textit{a priori} for the ``practical DMCT'' to
hold, if NP contributions in the neutrino interactions are allowed, as
in this case Eq.~(\ref{eq:general-D-M}) ``Direct'' and ``Exchange''
terms in general do not need to cancel each other.  To illustrate it
more clearly using symmetry properties of the transition amplitude,
let us assume that some (yet unknown) NP at high energy modifies the
low energy effective neutrino interactions we want to explore in
processes of the form $X \to Y \, \nu \, \overline{\nu}$, and the SM
singlet right-handed neutrinos and left-handed antineutrinos might as
well participate in such interactions.

\subsection{DMCT from the perspective of exchange symmetry}

Although process specific analysis using exchange symmetry is found in
the literature, a process-independent general analysis highlighting
the difference between Dirac and Majorana neutrinos would be useful in
context of NP interactions. Below we provide such an analysis.

The amplitude for $X(p_X) \to Y(p_Y) \, \nu(p_1) \,
\overline{\nu}(p_2)$ for Majorana neutrinos must be still
anti-symmetric with respect to the 4-momentum exchange $p_1
\leftrightarrow p_2$, as was noted in Eq.~\eqref{eq:amp-M}.  The Dirac
amplitude of Eq.~\eqref{eq:amp-D} could, in principle, be split into
two parts, one symmetric and the other anti-symmetric under the
exchange $p_1 \leftrightarrow p_2$, i.e.\ %
\begin{equation}
\label{eq:amp-D-sym-antisym}
\mathscr{M}^D = \mathscr{M}(p_1, p_2) = \mathscr{M}_\textrm{symm}(p_1,
p_2) + \mathscr{M}_\textrm{anti-symm}(p_1, p_2),
\end{equation}
where by definition we have
\begin{subequations}
\begin{align}
\mathscr{M}_\textrm{symm}(p_1, p_2) &= \frac{1}{2}
\Big(\mathscr{M}(p_1, p_2) + \mathscr{M}(p_2, p_1) \Big) \nonumber\\%
&=
\mathscr{M}_\textrm{symm}(p_2, p_1),
\label{eq:amp-symmetric}\\
\mathscr{M}_\textrm{anti-symm}(p_1, p_2) &= \frac{1}{2}
\Big(\mathscr{M}(p_1, p_2) - \mathscr{M}(p_2, p_1) \Big) \nonumber\\%
&=
-\,\mathscr{M}_\textrm{anti-symm}(p_2,
p_1),
\label{eq:amp-antisymmetric}
\end{align}
\end{subequations}
so that the Majorana amplitude of Eq.~\eqref{eq:amp-M} is
automatically given by,
\begin{equation}
\label{eq:amp-M-antisym}
\mathscr{M}^M = \sqrt{2} \, \mathscr{M}_\textrm{anti-symm}(p_1, p_2).
\end{equation}
With the decomposition of amplitudes as shown in
Eqs.~\eqref{eq:amp-D-sym-antisym} and \eqref{eq:amp-M-antisym}, the
difference between Dirac and Majorana neutrinos is given by,
\begin{align}
\left|\mathscr{M}^D\right|^2 -\left|\mathscr{M}^M\right|^2 &=
\left|\mathscr{M}_\textrm{symm}(p_1, p_2) +
\mathscr{M}_\textrm{anti-symm}(p_1, p_2)\right|^2 \nonumber\\%
&\quad - 2
\left|\mathscr{M}_\textrm{anti-symm}(p_1, p_2)\right|^2\,,\nonumber\\
&= \left|\mathscr{M}_\textrm{symm}(p_1, p_2)\right|^2 -
\left|\mathscr{M}_\textrm{anti-symm}(p_1, p_2)\right|^2 \nonumber\\
&\quad + 2\, \textrm{Re}\left(\mathscr{M}_\textrm{symm}(p_1, p_2)^* \,
\mathscr{M}_\textrm{anti-symm}(p_1, p_2)\right)\,.
\label{eq:D-M-diff}
\end{align}
If one were to do full integration over the neutrino and antineutrino
4-momenta, the interference term in Eq.~\eqref{eq:D-M-diff} vanishes
as it is anti-symmetric under the $p_1\leftrightarrow p_2$ exchange
and one is left with
\begin{align}
&\iint \left(\left|\mathscr{M}^D\right|^2 -\left|\mathscr{M}^M\right|^2
\right) \mathrm{d}^4p_1 \, \mathrm{d}^4 p_2 \nonumber\\%
=&\iint \left(
\left|\mathscr{M}_\textrm{symm}(p_1, p_2)\right|^2 -
\left|\mathscr{M}_\textrm{anti-symm}(p_1, p_2)\right|^2\right)
\mathrm{d}^4p_1 \, \mathrm{d}^4 p_2,
\label{eq:np-dmct}
\end{align}
which does not necessarily vanish in presence of NP interactions.
Note that Eqs.~\eqref{eq:integrated-D-M} and \eqref{eq:np-dmct} are
equivalent to one another, but Eq.~\eqref{eq:np-dmct} highlights the
symmetry properties that were not explicitly evident in
Eq.~\eqref{eq:integrated-D-M}.

\subsection{Processes with $\boldsymbol{\nu\,\overline{\nu}}$ produced via neutral current interactions}

As an example, let us consider NP effects on neutrino-antineutrino
neutral current interactions.  In such a case, the amplitude would
have the generic form,
\begin{equation*}
\mathscr{M}(p_1,p_2) = \sum_i \mathcal{S}_i \, \Big[ \overline{u}(p_1)
  \, \Gamma_i \, \varv(p_2) \Big],
\end{equation*}
where $\mathcal{S}_i$ denotes the parts of the amplitude which do not
depend on $p_1$ and $p_2$ or contain algebraic expressions such as
$(p_1 + p_2)^2$ which are \textit{symmetric} under $p_1
\leftrightarrow p_2$ exchange, and $\Gamma_i$ could be $\mathbf{1}$,
$\gamma^5$, $\gamma^\alpha$, $\gamma^\alpha \, \gamma^5$,
$\sigma^{\alpha\beta}$ or $\sigma^{\alpha\beta} \, \gamma^5$, which
respectively correspond to scalar, pseudo-scalar, vector,
axial-vector, tensor and axial-tensor interactions.  For Majorana
neutrinos it is well known that $\overline{u}(p_1) \, \Gamma \,
\varv(p_2) = - \, \overline{u}(p_2) \, C \, \Gamma^T \, C^{-1} \,
\varv(p_1)$ \cite{Denner:1992vza} where $C$ denotes the charge
conjugation operator.  Using this it is easy to show that
$\mathscr{M}_\textrm{symm}(p_1,p_2)$ gets contributions only from
vector, tensor and axial-tensor interactions while
$\mathscr{M}_\textrm{anti-symm}(p_1,p_2)$ gets contributions from
scalar, pseudo-scalar and axial-vector interactions alone, i.e.\ %
\begin{align*}
\mathscr{M} (p_1,p_2) &= \sum_i \mathcal{S}_i \, \Big[ \overline{u}(p_1) \,
\Gamma_i \, \varv(p_2) \Big] \\%
&= 
\begin{cases}
\mathscr{M}_\textrm{symm}(p_1,p_2) & \text{ for }
\Gamma_i=\gamma^\alpha, \, \sigma^{\alpha\beta}, \,
\sigma^{\alpha\beta} \,\gamma^5,\\ %
\mathscr{M}_\textrm{anti-symm}(p_1,p_2) & \text{ for } \Gamma_i=
\mathbf{1}, \, \gamma^5, \, \gamma^\alpha \, \gamma^5.
\end{cases}
\end{align*}
Therefore, the NP effects could lead to observable differences between
the Dirac and Majorana neutrino interactions, even after integrating
over the unknown neutrino and antineutrino 4-momenta, if the symmetric
and anti-symmetric parts of the amplitude do not cancel either due to
some imposed symmetry (like in the case of vector $V-A$ interactions
in the SM) or due to some accidental fine-tuning and numerical
cancellation between different types of couplings.  In
section~\ref{sec:examples} we illustrate the general considerations
presented above discussing a chosen physical process of the form $X
\to Y \, \nu \, \overline{\nu}$.

\section{Example 2-body and 3-body neutral current decays with $\boldsymbol{\nu\,\overline{\nu}}$ in the final state}
\label{sec:examples}

To illustrate the results arising from the general formalism presented
in previous sections, we study here in detail two example processes,
in both cases assuming the non-vanishing NP effects, namely the 2-body
decay of $Z^0$ boson, $Z \to\nu_\ell \, \overline{\nu}_\ell$ and the
3-body decay of an initial pseudo-scalar meson to a final
pseudo-scalar meson and $\nu\, \overline{\nu}$.  A recent
model-independent analysis of 3-body leptonic decays, but in context
of heavy sterile neutrinos and facilitated by charged current
interaction, can be found in Ref.~\cite{Marquez:2022bpg}.

\subsection{2-body decay of $\boldsymbol{Z \to \nu_\ell \, \overline{\nu}_\ell}$}
\label{sec:zvv}

Considering Lorentz invariance alone, it is possible to write down the
most general $Z(p) \to \nu(p_1) \, \overline{\nu}(p_2)$ decay
amplitude as follows (for Dirac neutrinos),
\begin{align}
\mathscr{M}^D &= -i\,\epsilon_\alpha(p) \, \overline{u}(p_1) \, \Big[
\Big( g_S^+ + g_P^+ \, \gamma^5 \Big) \, p^\alpha + \Big( g_S^- +
g_P^- \, \gamma^5 \Big) \, q^\alpha \nonumber\\*%
&\hspace{23mm} + \gamma^\alpha \Big( g_V + g_A \, \gamma^5 \Big) +
\sigma^{\alpha\beta} \Big(g_{T_\textrm{md}}^+ + g_{T_\textrm{ed}}^+
\gamma^5 \Big) \, p_\beta \nonumber\\*%
&\hspace{23mm} + \sigma^{\alpha\beta} \Big(g_{T_\textrm{md}}^- +
g_{T_\textrm{ed}}^- \gamma^5 \Big) \, q_\beta \Big]\,
\varv(p_2),\label{eq:Znn-amp-D} %
\end{align}
where $\epsilon_\alpha(p)$ denotes the polarisation 4-vector of $Z$,
$q=p_1 - p_2$, $p=p_1 + p_2$, $g_X^{(\pm)}$ with $X = S, P, V, A,
T_\textrm{md}, T_\textrm{ed}$ denote the various possible coupling
constants corresponding to scalar, pseudo-scalar, vector,
axial-vector, tensor (magnetic dipole) and axial-tensor (electric
dipole) kind of interactions.  In the SM, $g_S^\pm = g_P^\pm =
g_{T_\textrm{md}}^\pm = g_{T_\textrm{ed}}^\pm = 0$ and $g_V = - g_A =
\dfrac{g_Z}{4}$ where $g_Z = e/(\sin\theta_W\,\cos\theta_W)$ with
$\theta_W$ being the weak mixing angle and $e$ being the electric
charge of positron.  In presence of NP all the above coupling
constants could, in principle, differ from the SM values.  Noting that
all terms proportional to $p^\alpha$ vanish since $p^\alpha \,
\epsilon_\alpha (p) = 0$, and utilising Gordon identities as well as
neglecting terms proportional to $m_\nu$, the tensorial components get
eliminated to yield,
\begin{equation}
\mathscr{M}^D = -i\,\epsilon_\alpha(p) \, \overline{u}(p_1) \, \Big[
  \Big( g_S + g_P \, \gamma^5 \Big) \, q^\alpha + \gamma^\alpha \Big(
  g_V + g_A \, \gamma^5 \Big) \Big]\, \varv(p_2),\label{eq:Amp-D-1}
\end{equation}
where $g_S = g_S^- + i\,g_{T_\textrm{md}}^+$ and $g_P = g_P^- +
i\,g_{T_\textrm{ed}}^+$.  Further, if we consider CP and CPT
conservation in the decay $Z \to \nu_\ell \, \overline{\nu}_\ell$,
$g_S =0$ and $g_P=0$ ought to be satisfied respectively.  This
implies, only vector and axial-vector couplings are allowed.  The most
general expression for the decay amplitude for $Z \to \nu_\ell \,
\overline{\nu}_\ell$ with Dirac neutrinos is given by,
\begin{equation}
\label{eq:Z-nn-D-amp}
\mathscr{M}^D = \mathscr{M}(p_1, p_2) = - \frac{i\, g_Z}{2}
\upepsilon_\alpha(p) \, \left[\overline{u}(p_1) \,
  \gamma^\alpha\left(C_V^\ell-C_A^\ell\, \gamma^5\right) \,
  \varv(p_2)\right],
\end{equation}
where, to keep our analysis more general, we have elevated the
parameters $g_V$ and $g_A$ to possibly be different for every the
lepton family by adding the superscript $\ell=e,\mu,\tau$ such that
the vector and axial-vector couplings are expressed as $g_V^\ell =
\dfrac{g_Z}{2} C_V^\ell$ and $g_A^\ell = - \dfrac{g_Z}{2} C_A^\ell$
respectively, with %
\begin{equation}
C_{V, A}^\ell = \frac{1}{2} + \varepsilon_{V, A}^\ell,
\end{equation}
where $\varepsilon_V^\ell$, $\varepsilon_A^\ell$ parameterise the NP
effects, vanishing in the SM case.  The amplitude for Majorana case is
given by %
\begin{align}
\mathscr{M}^M &= \dfrac{1}{\sqrt{2}} \Big(\mathscr{M}(p_1, p_2) -
\mathscr{M}(p_2, p_1) \Big) \nonumber\\%
&= \frac{i\, g_Z\, C_A^\ell}{\sqrt{2}}
\upepsilon_\alpha(p) \, \left[\overline{u}(p_1) \, \gamma^\alpha \,
  \gamma^5 \, \varv(p_2)\right].
\label{eq:Z-nn-M-amp}
\end{align}
It is clear that we can combine the direct and exchange amplitudes in
this case and effectively redefine the vertex structure for
$Z\to\nu_\ell\, \overline{\nu}_\ell$ when Majorana neutrinos are
considered.

Keeping neutrino mass dependent terms in the amplitude squares, we get
different results for Dirac and Majorana neutrinos:
\begin{align}
\left|\mathscr{M}^D\right|^2 &= \frac{g_Z^2}{3}
\Bigg(\left((C_V^\ell)^2 + (C_A^\ell)^2\right) \left(m_Z^2 -
m_\nu^2\right) \nonumber\\*%
&\hspace{2cm} + 3 \left((C_V^\ell)^2 - (C_A^\ell)^2\right)
m_\nu^2\Bigg), \\
\left|\mathscr{M}^M\right|^2 &= \frac{2\, g_Z^2\, (C_A^\ell)^2}{3}
\left(m_Z^2 - 4\, m_\nu^2\right),
\end{align}
such that
\begin{align}
\left|\mathscr{M}^D\right|^2 - \left|\mathscr{M}^M\right|^2 &=
\frac{g_Z^2}{3} \bigg(\left((C_V^\ell)^2-(C_A^\ell)^2\right)\left(m_Z^2+2\,
m_\nu^2\right) \nonumber\\*%
&\hspace{1cm} + 6\, (C_A^\ell)^2\, m_\nu^2\bigg) \nonumber\\*
& =
\begin{cases}
\dfrac{g_Z^2}{2} m_\nu^2, & \left(\parbox[c]{2cm}{\centering for SM alone} \right)\\[5mm]
\dfrac{g_Z^2}{3} \left(\varepsilon_V^\ell-\varepsilon_A^\ell\right) m_Z^2, & \left(\parbox[c]{2cm}{\centering with NP but neglecting $m_\nu$}\right)
\end{cases}
\end{align}
where we have kept only the leading order contributions of
$\varepsilon_{V, A}^\ell$ while considering NP effects.  It is clear
that the SM result is fully in agreement with ``practical Dirac
Majorana confusion theorem''.  However, the difference between Dirac
and Majorana neutrinos appears in context of NP contributions even
when one neglects $m_\nu$ dependent terms (unless, of course,
$\varepsilon_V^\ell = \varepsilon_A^\ell$ in which case the additional
NP contributions effectively rescale the SM allowed $V-A$ coupling).
Possible example of NP effects in this $Z$ boson decay could arise
from kinetic mixing of $Z$ with the neutral gauge bosons from extra
gauge groups like additional $U(1)$ or $SU(2)_R$.  In this paper we
are not concerned with any specific model of NP to keep our results
very general.

\begin{figure*}[h!]
\centering %
\includegraphics[width=0.7\linewidth,keepaspectratio]{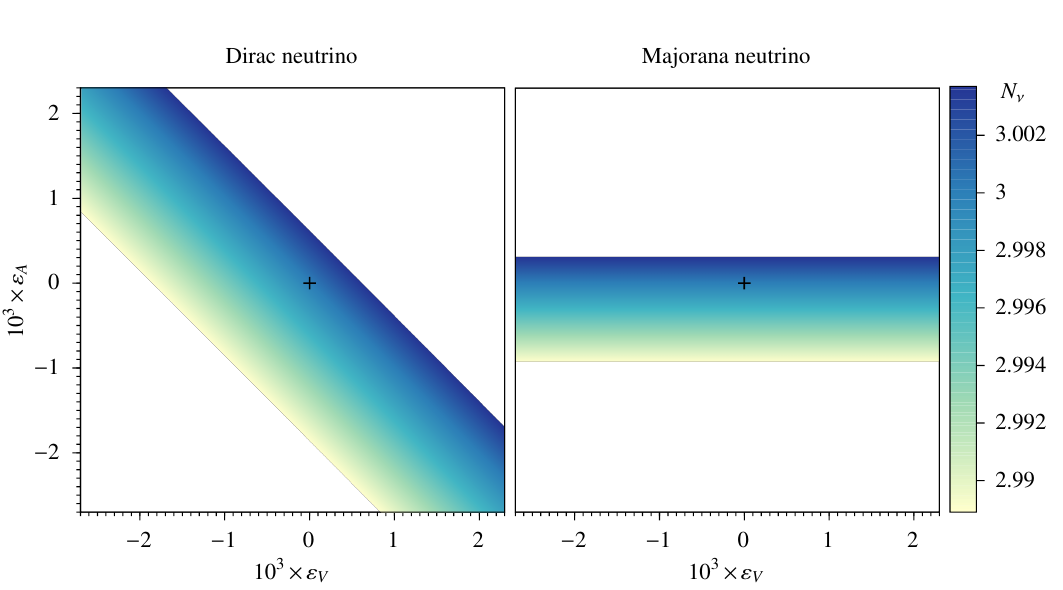} %
\caption{Constraints on the NP parameters $\varepsilon_V$ and
  $\varepsilon_A$ from the experimentally measured number of light
  neutrino species $N_\nu$.  The point denoted by `$+$' corresponds to
  the SM values $\varepsilon_V = 0 = \varepsilon_A$.  In the plots
  here $\varepsilon_{V,A}^2$ contributions have not been neglected.} %
\label{fig:Nnu}
\end{figure*}

The decay width of $Z$ boson into invisible final states is well
measured and can constrain NP contributions to neutrino couplings.
Neglecting small neutrino masses, the decay rates for $Z\to \nu_\ell
\, \overline{\nu}_\ell$ for Dirac and Majorana neutrino possibilities
are given by
\begin{subequations}
\begin{align}
\Gamma^D (Z\to \nu_\ell\, \overline{\nu}_\ell) &= \Gamma_Z^{0} \left(1
+ 2\, \varepsilon_V^\ell + 2\, \varepsilon_A^\ell \right),
\label{eq:Z2nnD-rate}\\
\Gamma^M (Z\to \nu_\ell\, \overline{\nu}_\ell) &= \Gamma_Z^{0} \left(1
+ 4\, \varepsilon_A^\ell\right),
\label{eq:Z2nnM-rate}
\end{align}
\end{subequations}
where $\Gamma_Z^{0}$ denotes the SM decay rate for $m_\nu=0$,
\begin{equation}
\Gamma_Z^{0} = \dfrac{G_F\, m_Z^3}{12\sqrt{2}\, \pi},
\end{equation}
with $G_F$ being the Fermi constant.  Thus, in presence of NP effects
and neglecting $m_\nu$ dependent terms, the total invisible width of
$Z$ boson is given by,
\begin{equation}
\label{eq:Z-inv-rate}
\Gamma_{Z, \textrm{inv}} =
\begin{cases}
\displaystyle \Gamma_Z^0 \, \Bigg(3 + 2 \sum_{\ell=e, \mu, \tau}
\left( \varepsilon_V^\ell + \varepsilon_A^\ell \right)\Bigg), &
\left(\parbox[c]{2cm}{\centering for Dirac neutrinos} \right)\\[5mm]
\displaystyle \Gamma_Z^0 \, \Bigg(3 + 4 \sum_{\ell=e, \mu, \tau}
\varepsilon_A^\ell \Bigg).   & \left(\parbox[c]{2cm}{\centering for Majorana neutrinos} \right)
\end{cases}
\end{equation}
The ratio $\Gamma_{Z, \textrm{inv}}/\Gamma_Z^0$ is a measure of the
number of light neutrino species $N_\nu$ and its experimental estimate
is \cite{ParticleDataGroup:2020ssz, Janot:2019oyi},
\begin{equation}\label{eq:Nnu}
N_\nu = \Gamma_{Z, \textrm{inv}}/\Gamma_Z^0 = 2.9963 \pm 0.0074.
\end{equation}
Using Eq.~\eqref{eq:Z-inv-rate} in Eq.~\eqref{eq:Nnu} we get the
following constraints on the NP parameters,
\begin{subequations}\label{eq:NP-constraints}
\begin{align}
\sum_{\ell=e, \mu, \tau} \left( \varepsilon_V^\ell +
\varepsilon_A^\ell \right) &= -0.0018 \pm 0.0037 , &&\left(\parbox[c]{2cm}{\centering for
  Dirac neutrinos} \right)\\
\sum_{\ell=e, \mu, \tau} \varepsilon_A^\ell &= -0.0009 \pm 0.0018,
&&\left(\parbox[c]{2cm}{\centering for Majorana neutrinos} \right)
\end{align}
\end{subequations}
which are perfectly consistent with zero, however obviously different
for Majorana and Dirac neutrinos.  In the latter case, it is in
principle even possible that deviations from the SM $V-A$-type
couplings are individually larger but cancel between the vector and
axial contributions.

Assuming that the NP contributions do not depend on the flavor of the
neutrino, i.e.\ $\varepsilon_{V,A}^e = \varepsilon_{V,A}^\mu =
\varepsilon_{V,A}^\tau \equiv \varepsilon_{V,A}$~(say), and keeping
the contributions from $\varepsilon_{V,A}^2$ terms as well, we find
that $\varepsilon_V$ and $\varepsilon_A$ get constrained by
Eq.~\eqref{eq:Nnu} as shown in Fig.~\ref{fig:Nnu}.  It is clear from
Fig.~\ref{fig:Nnu} that the SM values $\varepsilon_V = 0 =
\varepsilon_A$ (corresponding to the point indicated by `$+$') are
within the allowed $1\sigma$ region.

The results above can be interpreted in two-fold way. First, one
can assume some specific NP model that affects the weak neutral
current processes involving neutrinos, with well defined Lagrangian
and field interactions. In this case nature of the neutrinos follows
from the model construction.  As always, to predict the
phenomenological consequences of the given SM extension, one needs
first to estimate the experimentally allowed size of NP couplings. An
example above illustrates how such bounds differ for a particular
decay depending on the type of neutrinos in a model. Obviously, the
realistic NP models usually contain a number of new parameters and
constraining them requires combining bounds from various measurements,
often not only from the neutrino sector. The invisible $Z$ boson decay
discussed in this section, as a quantity known to a good accuracy,
should always be one of observables contributing to such more general
analysis and fitting NP model to data.

Second, it may be not known in advance what is the nature of the
neutrino fields, apart from the assumption that they can have
interactions going beyond the SM-like $V-A$ couplings.  In this case
single measurement like the invisible $Z$ boson decay cannot help
distinguish the type of neutrino fields if some deviations from the SM
predictions are observed (or even if neutrinos are responsible for
such deviations). However, again when more measurements are combined
(especially for the final states with more than 2 particles, which
allow to construct additional observables like shown in the next
section) it may be possible to constrain individual couplings and, in
an optimistic scenario, understand the nature of ``invisible'' sector
of the theory. Again, reconstructing the model couplings must take
into account the fact that it depends on the nature of neutrino
fields.

Finally, apart from specific decay discussed in this section,
the general formalism presented before can be applied to other 2-body
neutrino decays which can be considered in searches for various
channels of decays of BSM particles, e.g. new neutral heavy gauge
bosons or scalars (here possible example are the sneutrino decays in
R-parity violating MSSM).

\subsection{3-body decay of an initial pseudo-scalar meson to a
  final pseudo-scalar meson and $\boldsymbol{\nu\,\overline{\nu}}$}
\label{sec:3body}

As a second example let us consider the general decay $\mathcal{P}_i
\to \mathcal{P}_f \, \nu_\ell \, \overline{\nu}_\ell$ where
$\mathcal{P}_i$ is the parent pseudo-scalar meson with mass $M_i$ and
$\mathcal{P}_f$ is the daughter pseudo-scalar meson with mass $M_f$.
For example, $\mathcal{P}_i$ could be $B$ or $K$ meson, then
$\mathcal{P}_f$ would be either $K$ or $\pi$ meson respectively.

Considering Lorentz invariance, the effective Lagrangian for
$\mathcal{P}_i \to \mathcal{P}_f \, \nu_\ell \, \overline{\nu}_\ell$
decay can be written as follows,
\begin{align}
\mathscr{L} &= J_{SL}^\ell \Big(\overline{\psi}_\nu \,P_L\,
\psi_{\overline{\nu}} \Big) + J_{SR}^\ell \Big(\overline{\psi}_\nu \, P_R
\, \psi_{\overline{\nu}} \Big) \nonumber\\*%
&\quad + \left(J_{VL}^\ell\right)_{\alpha}
\Big(\overline{\psi}_\nu \, \gamma^\alpha P_L \, \psi_{\overline{\nu}}
\Big) + \left(J_{VR}^\ell\right)_\alpha \Big(\overline{\psi}_\nu \,
\gamma^\alpha \, P_R \, \psi_{\overline{\nu}} \Big) \nonumber\\*
&\quad + \left(J_{TL}^\ell\right)_{\alpha\beta} \Big(\overline{\psi}_\nu
\,P_L\, \sigma^{\alpha\beta} \, \psi_{\overline{\nu}} \Big) +
\left(J_{TR}^\ell\right)_{\alpha\beta} \Big(\overline{\psi}_\nu \,
\sigma^{\alpha\beta}\, P_R \, \psi_{\overline{\nu}} \Big) +
\textrm{h.c.},
\label{eq:effective-Lagrangian}
\end{align}
where $\psi_j$ denotes the fermionic field of $j=\nu_\ell,
\overline{\nu}_\ell$ and $J_{SL}^\ell$, $J_{SR}^\ell$,
$\left(J_{VL}^\ell\right)_\alpha$, $\left(J_{VR}^\ell \right)_\alpha$,
$\left(J_{TL}^\ell\right)_{\alpha\beta}$
$\left(J_{TR}^\ell\right)_{\alpha\beta}$ denote the different hadronic
currents describing the quark level transitions from $\mathcal{P}_i$
to $\mathcal{P}_f$ meson.  We have used the superscript $\ell$ to
accommodate any possible differences in the effective Lagrangian when
different neutrino flavors are considered.  The subscripts $S$, $V$
and $T$ in the hadronic currents denote the fact that the associated
external leptonic currents are scalar, vector and tensor type,
respectively, and in addition they carry also the chirality index.  In
the SM, at tree-level, only the $V-A$ interaction is present, while
the scalar and tensor interactions involving left-handed neutrinos
might be generated from quantum loops and are hence expected to be
suppressed.  Additionally, in the SM the right-handed neutrinos do not
take part in any interaction.  In this analysis we do not consider any
specific NP model to keep our results general, thus in
Eq.~\eqref{eq:effective-Lagrangian} we have included all possible
interactions disregarding the helicity of neutrino.

The decay amplitude for $\mathcal{P}_i (p_i) \to \mathcal{P}_f (p_f)
\, \nu_\ell (p_1) \, \overline{\nu}_\ell (p_2)$, considering the
neutrino and the antineutrino to be Dirac fermions, is given by,
\begin{align}
\mathscr{M}^D = \mathscr{M}(p_1, p_2) &= \overline{u}(p_1) \Big[
F_{SL}^\ell \, P_L\, + F_{SR}^\ell \, P_R \nonumber\\*%
&\hspace{15mm} + \left(F_{VL}^{\ell +} \, p_\alpha + F_{VL}^{\ell -}
q_\alpha \right) \, \gamma^\alpha\, P_L \nonumber\\*%
&\hspace{15mm} + \left(F_{VR}^{\ell +} \, p_\alpha + F_{VR}^{\ell -}
q_\alpha \right) \, \gamma^\alpha \, P_R \, \nonumber\\*%
&\hspace{15mm} + F_{TL}^\ell \, p_\alpha \, q_\beta \,
\sigma^{\alpha\beta} \,P_L \nonumber\\*%
&\hspace{15mm} + F_{TR}^\ell \, p_\alpha \, q_\beta \,
\sigma^{\alpha\beta} \, P_R \Big] \varv(p_2), %
\label{eq:D-amplitude} %
\end{align}
where $p=p_i + p_f$, $q=p_i-p_f = p_1 + p_2$ and the various form
factors $F_{SX}^\ell$, $F_{VX}^{\ell \pm}$, $F_{TX}^\ell$ are defined
as follows,
\begin{subequations}\label{eq:form-factors}
\begin{align}
\left\langle \mathcal{P}_f \Big\lvert J_{SX}^\ell \Big\rvert
\mathcal{P}_i \right\rangle &= F_{SX}^\ell, \\
\left\langle \mathcal{P}_f \Big\lvert \left(J_{VX}^\ell\right)_\alpha
\Big\rvert \mathcal{P}_i \right\rangle &= F_{VX}^{\ell +} \, p_\alpha
+ F_{VX}^{\ell -} \, q_\alpha, \\
\left\langle \mathcal{P}_f \Big\lvert
\left(J_{TX}^\ell\right)_{\alpha\beta} \Big\rvert \mathcal{P}_i
\right\rangle &= F_{TX}^\ell \, p_\alpha \, q_\beta,
\end{align}
\end{subequations}
where by $X$ we denote the chirality index, $X=L,R$.

The form factors defined in Eq.~\eqref{eq:form-factors} are complex
and, in general, functions of the square of the momentum transferred,
i.e.\ $q^2\equiv s$.  Here the form factors also include the CKM
matrix elements and their mass dimensions are different since the
explicit dependencies on quark/meson masses are hidden for simplicity
of expressions.

The decay amplitude for Majorana neutrinos is obtained by
anti-symmetrization with respect to exchange of $p_1$ and $p_2$ and is
given by,
\begin{align}
\mathscr{M}^M &= \frac{1}{\sqrt{2}} \Big(\mathscr{M}(p_1, p_2) -
\mathscr{M}(p_2, p_1)\Big) \nonumber\\*
&= \sqrt{2} \, \overline{u}(p_1) \,\Bigg[ F_{SL}^\ell \,P_L +
  F_{SR}^\ell \, P_R \nonumber\\%
&\hspace{16mm} + \bigg(\frac{F_{VR}^{\ell +} - F_{VL}^{\ell
      +}}{2} \, p_\alpha + \frac{F_{VR}^{\ell -} - F_{VL}^{\ell -}}{2}
  q_\alpha \bigg) \, \gamma^\alpha \, \gamma^5 \Bigg]\, \varv(p_2).
\label{eq:M-amplitude}
\end{align}
Once again we find that the direct and exchange amplitudes in this
case can be combined to effectively redefine the vertex factors for
each possible interaction.  Notably, for Majorana neutrinos no vector
or tensor neutral currents are possible.

\begin{figure}[hbtp]
\centering
\includegraphics[width=0.8\linewidth,keepaspectratio]{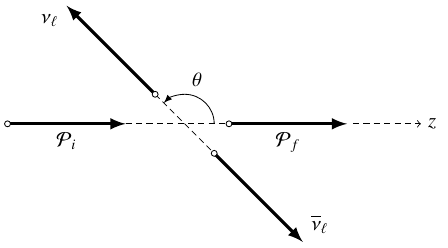}
\caption{Kinematic configuration of the decay $\mathcal{P}_i \to
  \mathcal{P}_f \, \nu \, \overline{\nu}$ in center-of-momentum frame
  of the $\nu\, \overline{\nu}$ pair.}
\label{fig:com}
\end{figure}

Since the scalar products that appear in the amplitude squares are
Lorentz invariant, they can be evaluated in any chosen frame of
reference.  It is well known that any 3-body decay can be fully
described by two independent parameters.  For simplicity we choose
them as $s = \big(p_1 + p_2\big)^2$ and the angle $\theta$ between the
3-momenta of the neutrino and the final state meson in the
center-of-momentum frame of the $\nu\, \overline{\nu}$ pair, as shown
in Fig.~\ref{fig:com}.

The amplitude squares for both Dirac and Majorana cases can be written
in the following form in terms of the variables $s$ and $\cos\theta$,
\begin{equation}\label{eq:AmpSq-DM}
\left\lvert \mathscr{M}^{D/M} \right\rvert^2 = C_0^{D/M} + C_1^{D/M} \,
\cos\theta + C_2^{D/M} \,  \cos^2\theta,
\end{equation}
where, neglecting the small terms proportional to neutrino mass
$m_\nu$ we have,
\begin{subequations}
\begin{align}
C_0^D &= s\,\Big( \modulus{F_{SL}^\ell}^2 + \modulus{F_{SR}^\ell}^2 \Big) +
\lambda \, \Big(\modulus{F_{VL}^{\ell +}}^2 + \modulus{F_{VR}^{\ell +}}^2\Big) , \\
C_1^D &= 2\, s \,\sqrt{\lambda}\, \left(\Im\left(F_{SL}^\ell
F_{TL}^{\ell *}\right) + \Im\left(F_{SR}^\ell F_{TR}^{\ell *} \right)\right)
, \label{eq:C1D}\\
C_2^D &= - \lambda \, \Big( \modulus{F_{VL}^{\ell +}}^2 +
\modulus{F_{VR}^{\ell +}}^2 - s \, \left( \modulus{F_{TL}^\ell}^2 +
\modulus{F_{TR}^\ell}^2 \right) \Big), \\[3mm]
C_0^M &= 2 \,s\, \Big(\modulus{F_{SL}^\ell}^2 + \modulus{F_{SR}^\ell}^2\Big) +
\lambda\, \modulus{F_{VL}^{\ell +} - F_{VR}^{\ell +}}^2 ,\\
C_1^M &= 0, \\
C_2^M &= - \lambda \, \modulus{F_{VL}^{\ell +} - F_{VR}^{\ell +}}^2 ,
\end{align}
\end{subequations}
with
\begin{equation}
\lambda = M_i^4 + M_f^4 + s^2 - 2 \left(M_i^2 \, M_f^2 + s \, M_i^2 +
s \, M_f^2\right).
\end{equation}
The Dirac amplitude square may contain a term odd in $\cos\theta$,
while it is always absent in Majorana case.  This, in principle, could
be probed if the angle $\theta$ could be measured experimentally.
However, at present the angle $\theta$ can not be observed as both the
neutrino and antineutrino remain undetected in the detector near their
point of production.  In addition, such a term also requires
simultaneously existence of both tensor and scalar NP interactions (in
addition, with complex couplings) and is likely to be small.

The differential decay rate in the rest frame of the parent meson
$\mathcal{P}_i$ after integration over the unobservable $\cos\theta$
is given by,
\begin{equation}\label{eq:dG-by-ds-in-rest-frame}
\frac{\mathrm{d}\Gamma^{D/M}}{\mathrm{d}s} = \frac{1}{(2\, \pi)^3}
\frac{b}{16\, M_i^3} \left(C_0^{D/M} + \frac{1}{3} C_2^{D/M}\right),
\end{equation}
where
\begin{equation}\label{eq:b}
b = \frac{\sqrt{\lambda}}{2} \, \sqrt{1- \frac{4\,m_\nu^2}{s}}.
\end{equation}

In terms of the form factors, we get
\begin{subequations}\label{eq:DistsDM}
\begin{align}
\frac{\mathrm{d}\Gamma^{D}}{\mathrm{d}s} &= \frac{1}{(2\, \pi)^3}
\frac{\lambda\,b}{24\, M_i^3} \Bigg( \modulus{F_{VL}^{\ell +}}^2 +
\modulus{F_{VR}^{\ell +}}^2 \nonumber\\%
&\hspace{25mm} + \frac{3s}{2\lambda} \, \left(
\modulus{F_{SL}^\ell}^2 + \modulus{F_{SR}^\ell}^2 \right) \nonumber\\%
&\hspace{25mm} +
\frac{s}{2} \, \left( \modulus{F_{TL}^\ell}^2 +
\modulus{F_{TR}^\ell}^2 \right)\Bigg)\, ,
\label{eq:D-dist}\\
\frac{\mathrm{d}\Gamma^{M}}{\mathrm{d}s} &= \frac{1}{(2\, \pi)^3}
\frac{\lambda \,b}{24\, M_i^3} \Bigg( \modulus{F_{VL}^{\ell +} -
  F_{VR}^{\ell +}}^2 + \frac{3s}{\lambda} \, \left(
\modulus{F_{SL}^\ell}^2 + \modulus{F_{SR}^\ell}^2 \right) \Bigg).
\label{eq:M-dist}
\end{align}
\end{subequations}
From Eqs.~\eqref{eq:D-dist} and \eqref{eq:M-dist} it is clear that the
presence of additional forms of NP interactions on top of the usual SM
allowed $V-A$ interaction can lead to observable differences between
the Dirac and Majorana neutrinos.

To show the impact of presence of such additional non-standard
interactions we consider a simple effective parametrization of NP
effects (using again the chirality index $X=L,R$),
\begin{subequations}\label{eq:NP-parameters}
\begin{align}
F_{SX}^\ell &= M_i \, F_{\SM} \, \varepsilon_{SX}^\ell, \\
F_{VL}^{\ell +} & = F_{\SM} \,\Big(1+\varepsilon_{VL}^\ell\Big), \\
F_{VR}^{\ell +} & = F_{\SM}\, \varepsilon_{VR}^\ell, \\
F_{TX}^\ell &= \frac{1}{M_i} \, F_{\SM} \, \varepsilon_{TX}^\ell,
\end{align}
\end{subequations}
where $\varepsilon_{QX}^\ell$, $Q=S,V,T$, denote the relative size of
the NP contributions with respect to the SM contribution\footnote{Note
that $\varepsilon_{QX}^\ell$ are in general different quantities than
$\varepsilon_V^\ell,\varepsilon_A^\ell$ defined in
section~\ref{sec:zvv}.} and are assumed to be small, $
\modulus{\varepsilon_{QX}^\ell} \ll 1$.  We include $M_i$ factors in
Eq.~(\ref{eq:NP-parameters}) to accommodate for the difference in mass
dimensions of various form factors so as to keep the NP parameters
$\varepsilon_{QX}^\ell$ dimensionless.  Substituting
Eq.~\eqref{eq:NP-parameters} in Eq.~\eqref{eq:DistsDM} we obtain, %
\begin{subequations}\label{eq:DM-dist-NP}
\begin{align}
\frac{\mathrm{d}\Gamma^{D}}{\mathrm{d}s} &=
\frac{\mathrm{d}\Gamma^{\SM}}{\mathrm{d}s} \Bigg(1+ 2\, \Re\,
\varepsilon_{VL}^\ell + \modulus{\varepsilon_{VL}^\ell}^2 +
\modulus{\varepsilon_{VR}^\ell}^2 \nonumber\\%
&\hspace{16mm} + \frac{3 s M_i^2}{2\lambda} \,
\left( \modulus{\varepsilon_{SL}^\ell}^2 +
\modulus{\varepsilon_{SR}^\ell}^2 \right) \nonumber\\%
&\hspace{16mm} + \frac{s}{2 M_i^2} \,
\left( \modulus{\varepsilon_{TL}^\ell}^2 +
\modulus{\varepsilon_{TR}^\ell}^2 \right)\Bigg) \nonumber\\*
&\approx \frac{\mathrm{d}\Gamma^{\SM}}{\mathrm{d}s} \Bigg(1+ 2\, \Re\,
\varepsilon_{VL}^\ell \Bigg)\, ,
\label{eq:D-dist-NP}\\
\frac{\mathrm{d}\Gamma^{M}}{\mathrm{d}s} &=
\frac{\mathrm{d}\Gamma^{\SM}}{\mathrm{d}s} \Bigg(1 + 2\, \Re\,
\varepsilon_{VL}^\ell - 2\, \Re\, \varepsilon_{VR}^\ell +
\modulus{\varepsilon_{VL}^\ell - \varepsilon_{VR}^\ell }^2 \nonumber\\%
&\hspace{16mm} + \frac{3 s
  M_i^2}{\lambda} \, \left( \modulus{\varepsilon_{SL}^\ell}^2 +
\modulus{\varepsilon_{SR}^\ell}^2 \right) \Bigg) \nonumber\\*
&\approx \frac{\mathrm{d}\Gamma^{\SM}}{\mathrm{d}s} \Bigg(1 + 2\,
\Re\, \varepsilon_{VL}^\ell - 2\, \Re\, \varepsilon_{VR}^\ell \Bigg),
\label{eq:M-dist-NP}
\end{align}
\end{subequations}
where we defined the SM decay rate as
\begin{equation}
\frac{\mathrm{d}\Gamma^{\SM}}{\mathrm{d}s} = \frac{1}{(2\, \pi)^3}
\frac{\lambda \, b\,\modulus{F_{\SM}}^2}{24\, M_i^3}\, .
\label{eq:gamsm}
\end{equation}
Therefore, the difference between Dirac and Majorana cases, to the
leading order in the NP parameters, is given by,
\begin{equation}
\frac{\mathrm{d}\Gamma^D}{\mathrm{d}s} -
\frac{\mathrm{d}\Gamma^M}{\mathrm{d}s} \approx 2 \,
\frac{\mathrm{d}\Gamma^{\SM}}{\mathrm{d}s} \,
\Re\,\varepsilon_{VR}^\ell,
\end{equation}
which implies that non-zero difference can, in principle, arise only
if $\Re\,\varepsilon_{VR}^\ell \neq 0$.

Similar to the case of $Z\to\nu\,\overline{\nu}$ decay, only the NP
contributions to the vector currents contribute to both decay rates in
the lowest linear order.  In addition, presence of only
$\varepsilon_{VL}$, i.e.  change of normalisation of the SM $V-A$
interaction, modifies the decay into Dirac neutrinos, while both
$\varepsilon_{VL}$ and $\varepsilon_{VR}$ affect the expression for
the Majorana neutrino decay.  Scalar and tensor interactions lead to
higher order corrections, quadratic in NP effects, and are likely to
be small and difficult to observe taking into experimental accuracy of
corresponding measurement, both current and in foreseeable future.
Moreover, if only constant vector or scalar NP contributions are
present, the shape of the decay spectrum
$\mathrm{d}\Gamma^{D/M}/\mathrm{d}s$ is identical for Majorana and
Dirac neutrinos and, therefore, when used alone it can not distinguish
between the two possibilities (unless in some specific NP model both
$\varepsilon_{VL}$ and $\varepsilon_{VR}$ are known and thus the
difference in normalisation of the spectrum is also known).  However,
it is interesting to note that if tensor NP contributions are
substantial enough to be observed, it would point out to the Dirac
nature of neutrino, since tensor interactions do not affect the
$s$ distribution for Majorana neutrinos at all.  We elaborate on these
aspects in Sec.~\ref{sec:dmdiff} where we study the $s$ distributions
while considering each of the possible interactions one at a time.

\begin{figure*}[hbtp]
\centering %
\includegraphics[width=0.7\linewidth,keepaspectratio]{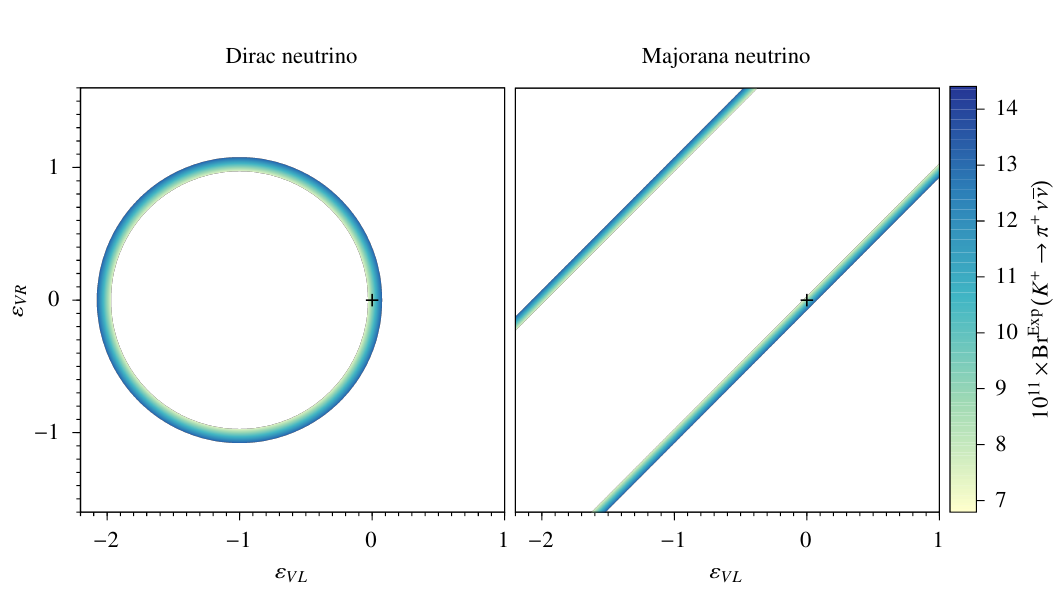} 
\caption{Constraints on the NP parameters $\varepsilon_{VL}$ and
  $\varepsilon_{VR}$ (assuming them to be real and including quadratic
  term contributions) using experimentally measured branching ratio of
  $K^+ \to \pi^+ \, \nu \, \overline{\nu}$ as given in
  Eq.~\eqref{eq:Br-exp}.  The point denoted by `$+$' corresponds to
  the SM values $\varepsilon_{VL} = 0 = \varepsilon_{VR}$.} %
\label{fig:epsilon-VLVR-constraint}
\end{figure*}

As an example of experimental constraints which can be imposed on the
NP neutrino couplings by the three-body decays, let us consider the
decay $K^+ \to \pi^+ \, \nu \, \overline{\nu}$ whose branching ratio
as reported by NA62 Collaboration \cite{NA62:2021zjw} is, %
\begin{equation}\label{eq:Br-exp}
\textrm{Br}^{\textrm{Exp}}\left(K^+ \to \pi^+ \, \nu \,
\overline{\nu}\right) =
\left(\left.   10.6^{+4.0}_{-3.4}\right|_\textrm{stat} \pm
0.9_{\textrm{syst}}\right) \times 10^{-11},
\end{equation}
which is consistent with the SM prediction \cite{Buras:2015qea},
\begin{equation}
\textrm{Br}^{\textrm{SM}}\left(K^+ \to \pi^+ \, \nu \,
\overline{\nu}\right) = \left(9.11 \pm 0.72\right) \times 10^{-11}.
\end{equation}
Neglecting the terms quadratic in NP contributions and assuming that
terms that are linear in the parameters $\varepsilon_{QX}$ are
approximately constant with respect to $q^2\equiv s$ (which could
affect the integration of $\mathrm{d}\Gamma/\mathrm{d}s$ to the total
branching ratio), we get the following constraints from the above
mentioned experimental measurement,
\begin{subequations}
\begin{align}
\sum_{\ell=e, \mu, \tau} \Re\,\varepsilon_{VL}^\ell &= 0.082 \pm
0.214, \quad \left( \parbox[c]{2cm}{\centering for Dirac neutrinos}\right)\\
\sum_{\ell=e, \mu, \tau} \left( \Re\,\varepsilon_{VL}^\ell -
\Re\,\varepsilon_{VR}^\ell \right) &= 0.082 \pm 0.214, \quad
\left( \parbox[c]{2cm}{\centering for Majorana neutrinos}\right)
\end{align}
\label{eq:nplimkll}
\end{subequations}
which are consistent with zero.

If we assume that only vector NP contributions are dominant and
$\varepsilon_{VL}^\ell$, $\varepsilon_{VR}^\ell$ are both real and
identical for all neutrino flavors, i.e.\ $\varepsilon_{VX}^e =
\varepsilon_{VX}^\mu = \varepsilon_{VX}^\tau \equiv \varepsilon_{VX}$
(with $X=L,R$), then the experimental measurement of
Eq.~\eqref{eq:Br-exp} constrains the NP parameters $\varepsilon_{VL}$
and $\varepsilon_{VR}$ as shown in
Fig.~\ref{fig:epsilon-VLVR-constraint} (there we have kept the
quadratic terms in $\varepsilon_{VL}$, $\varepsilon_{VR}$ as well,
assuming that in principle they can be large compared to the SM terms
but cancel out due to some kind of accidental or symmetry-related
fine-tuning).  It is clear from Fig.~\ref{fig:epsilon-VLVR-constraint}
that if we allow for some amount of fine tuning and cancellations,
large NP contributions are still possible in both Dirac and Majorana
neutrino possibilities.  One can also note that the contributions from
scalar and tensor form factors are always positive, so even if they
are also non-negligible, they can only tighten the bounds on
$\varepsilon_{VL}$ and $\varepsilon_{VR}$ plotted in
Fig.~\ref{fig:epsilon-VLVR-constraint}.

With more precise experimental decays data in future, the estimate of
NP contribution will also improve.  Another interesting decay in the
same category as $K^+\to\pi^+\,\nu\,\overline{\nu}$ is the decay $B^+
\to K^+ \, \nu \, \overline{\nu}$ which is being pursued at Belle~II
and awaiting its branching ratio measurement.  The SM prediction for
the branching ratio for this decay \cite{Blake:2016olu} is,
\begin{equation}
\textrm{Br}^{\SM} \left( B^+ \to K^+ \, \nu \, \overline{\nu} \right)
= \left(4.6 \pm 0.5 \right) \times 10^{-6},
\end{equation}
while the current experimental upper limit by Belle~II
\cite{Belle-II:2021rof} is $4.1 \times 10^{-5}$ at $90\%$ confidence
level.

Again, as already discussed in the previous section,
constraining individual parameters of the NP model (or in the best
case scenario even determining the nature of invisible sector of the
theory, including distinguishing the Dirac and Majorana character of
neutrinos) requires combining bounds from many measurements, in
addition to the example discussed above.

\section{Distinguishing Dirac and Majorana neutrinos via NP effects}
\label{sec:dmdiff}

Although experimentally measured branching ratios of the two- and
three-body decays can only constrain the probable NP couplings of
neutrinos, from Eq.~\eqref{eq:DistsDM} and the non-approximated part
of Eq.~\eqref{eq:DM-dist-NP} it seems that by analysing observed $s$
(missing mass-square) distribution in the three-body decays one can
also distinguish between Dirac and Majorana neutrino possibilities,
provided it can be measured with sufficient accuracy.  This could be
very challenging and may not possible in the near future as meson
decays into neutrino pair are rare and current experimental statistics
are not sufficient to probe the energy spectra of the visible final
state particles (equivalent to the $s$ distribution).  However, it
might eventually be possible when more data get accumulated.
Alternatively, similar analysis could be applied to collision
experiments with the neutrino antineutrino pair in the final state,
where statistics are much higher but one has to deal with significant
backgrounds and related uncertainties.

As mentioned in the Introduction, one should note that modification of
the spectra of the visible particle in the final state of the
decay can also be caused if some invisible particles other than
neutrinos are present in the final state. For distinguishing such NP
possibilities, i.e.\ the presence of other invisible particles, we
encourage the reader to follow other existing proposals such as the
ones discussed in Refs.~\cite{Goudzovski:2022vbt,Kim:2018hlp}.
Specifically, in context of 3-body semi-hadronic decays of mesons such
as $K \to \pi \, f \, \overline{f}$ where $f$ could be fermionic dark
matter, heavy neutral lepton or some long-lived fermion in addition to
the usual active neutrino possibility, the methods proposed in
Ref.~\cite{Kim:2018hlp} using angular distributions and Dalitz plot
asymmetries could be useful in distinguishing these NP possibilities,
without worrying about hadronic uncertainties. In the discussion
below, we focus only on active neutrinos as the invisible particles
and study the effect of NP interactions over the experimentally
observable distribution.%

Before proceeding further we note that in Eq.~\eqref{eq:DM-dist-NP},
$\varepsilon_{SL}$ and $\varepsilon_{SR}$ as well as
$\varepsilon_{TL}$ and $\varepsilon_{TR}$ contribute in a similar
manner.  Therefore, we can discuss their impact in terms of two
effective NP parameters $\varepsilon_S$ and $\varepsilon_T$, defined
as follows,
\begin{subequations}\label{eq:ST-NP-parameters}
\begin{align}
\varepsilon_S &= \sqrt{|\varepsilon_{SL}|^2 +|\varepsilon_{SR}|^2},\\
\varepsilon_T &= \sqrt{|\varepsilon_{TL}|^2 +|\varepsilon_{TR}|^2}.
\end{align}
\end{subequations}
Further we also consider, for simplicity, that all the NP parameters
are real quantities and agnostic to the flavor of the neutrino in the
final state.  Below we shall first look at the generic trends in the
various $s$ distributions for various NP possibilities without
considering any specific decay, and then we show how these generic
features can affect the spectrum of a specific decay, on an example
chosen to be $B \to K \, \nu \, \overline{\nu}$.

\subsection{The generic trends in $\boldsymbol{s}$ (missing mass-square)
  distributions}
\label{subsec:generic-trends}

\begin{figure*}[hbtp]
\centering%
\includegraphics[width=0.8\linewidth,keepaspectratio]{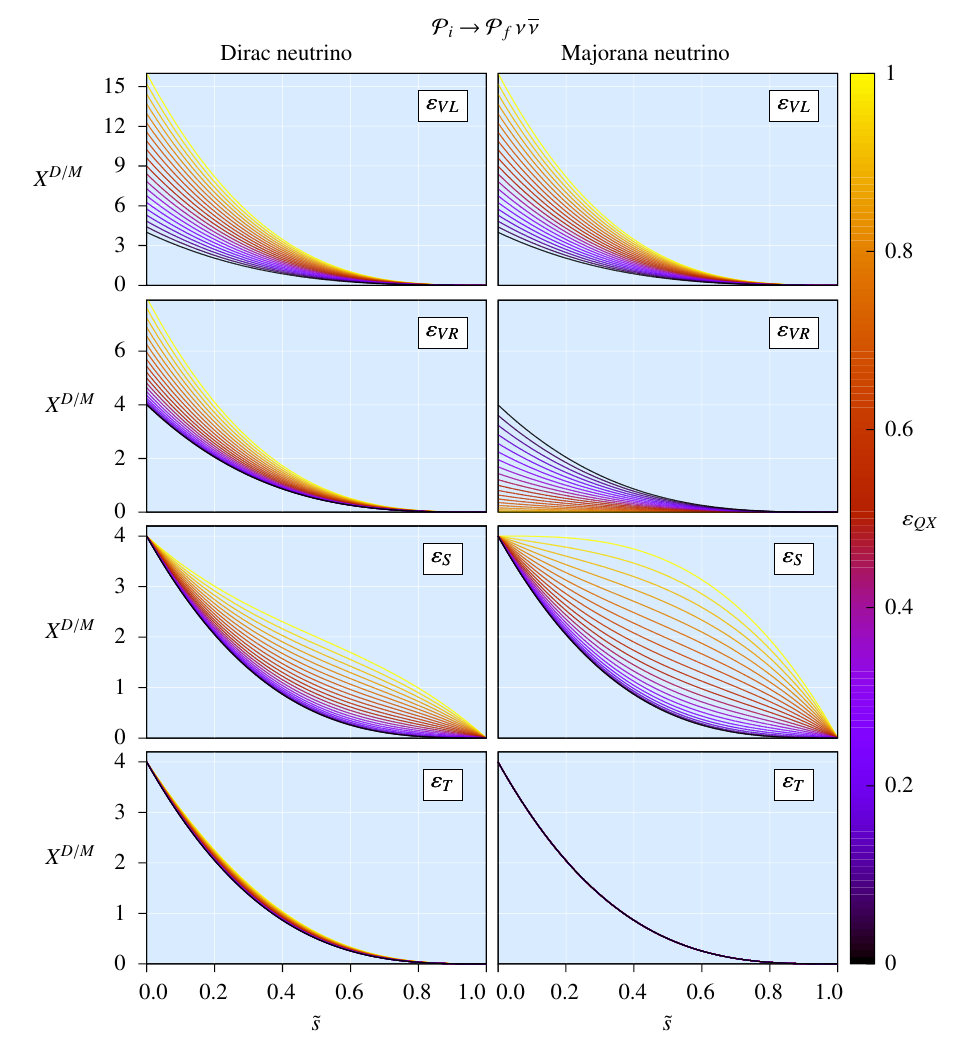}%
\caption{Comparison of various generic NP contributions in
  $\mathcal{P}_i \to \mathcal{P}_f \, \nu_\ell \, \overline{\nu}_\ell$
  decay assuming that the mass of the final meson $\mathcal{P}_f$ can
  be neglected in comparison with the mass of the initial meson
  $\mathcal{P}_i$ and assuming constant form factors.  As noted in the
  main text, all NP possibilities, except the left-handed vector NP
  contribution, can \textit{in principle} distinguish between Dirac
  and Majorana neutrinos.}%
\label{fig:Pi2Pfnn-NP}
\end{figure*}

It is possible to infer the overall trends and order of magnitude
effects of the NP contributions in the $s$ or missing mass-square
distributions, without considering any specific $\mathcal{P}_i \to
\mathcal{P}_f \, \nu \, \overline{\nu}$ decay, and by imposing the
following simplifying assumptions in Eqs.~\eqref{eq:D-dist-NP} and
\eqref{eq:M-dist-NP}:
\begin{enumerate}
\item the form factor $F_{\SM}$ has no $s$ dependence,
\item neglect $\mathcal{O}\left(M_f^2/M_i^2\right)$ terms, and
\item neglect the $m_\nu$ dependent term.
\end{enumerate}
It is clear that except the last assumption, the other two need not
even be good approximations in a real world example.  We will relax
these assumptions later while considering a specific decay.  For the
time being, we note that the above assumptions simplify the analysis
of Eq.~\eqref{eq:DM-dist-NP} which can now be cast into the discussion
of the dimensionless and normalised distributions $X^{D/M}$ defined
as, %
\begin{equation}
X^{D/M} = \frac{M_i^2}{\Gamma^{\SM}}
\frac{\mathrm{d}\Gamma^{D/M}}{\mathrm{d}s} = \frac{1}{\Gamma^{\SM}}
\frac{\mathrm{d}\Gamma^{D/M}}{\mathrm{d}\tilde{s}},
\end{equation}
with,
\begin{equation}
\Gamma^{\SM} \approx \frac{1}{(2\, \pi)^3}
\frac{M_i^5\,\modulus{F_{\SM}}^2}{192}\, .
\label{eq:gamtot}
\end{equation}
Using the simplifying assumptions, $X^{D/M}$ can be rewritten in the
following universal form,
\begin{subequations}\label{eq:XD-NP}
\begin{align}
X^{D} &= 4 \left(1-\tilde{s}\right) \Bigg(\left(1-\tilde{s}\right)^2
\left( 1+ 2\, \varepsilon_{VL} + \varepsilon_{VL}^2 +
\varepsilon_{VR}^2 + \frac{\tilde{s}}{2} \, \varepsilon_{T}^2 \right) \nonumber\\*%
&\hspace{2cm} + \frac{3}{2} \tilde{s} \, \varepsilon_{S}^2 \Bigg),
\label{eq:XD}\\
X^{M} &= 4 \left(1-\tilde{s}\right) \Bigg(\left(1-\tilde{s}\right)^2
\left( 1 + 2\, \varepsilon_{VL} - 2\, \varepsilon_{VR} +
\left(\varepsilon_{VL} - \varepsilon_{VR} \right)^2 \right) \nonumber\\*%
&\hspace{2cm} +
3\,\tilde{s} \, \varepsilon_{S}^2 \Bigg) ,
\label{eq:XM}
\end{align}
\end{subequations}
where $\tilde{s}=s/M_i^2$ is a dimensionless quantity which varies in
the range $[0,1]$.

To illustrate the order of the expected magnitude of effects of NP
neutrino couplings in the decay distributions, let us consider
contribution from each of the NP parameters individually and one at a
time.  In Fig.~\ref{fig:Pi2Pfnn-NP} we plot, in presence of individual
NP contributions, the distributions of the quantities $X^{D/M}$.  The
main features of these distributions are as follows.
\begin{enumerate}

\item The effect of left-handed vector NP contributions is identical
  for both Dirac and Majorana neutrinos.  Therefore, such a NP
  contribution would not lead to any difference between the Dirac and
  Majorana possibilities.

\item The right-handed vector NP contribution has opposite effects on
  Dirac and Majorana neutrinos.  For Dirac case the differential decay
  rate is enhanced, while for Majorana case it is reduced.  This
  opposite behaviour can clearly distinguish between the two
  possibilities.

\item As is also clear from Eqs.~\eqref{eq:XD} and \eqref{eq:XM} (also
  Eqs.~\eqref{eq:D-dist-NP} and \eqref{eq:M-dist-NP}), the effect
  of scalar NP contributions is enhanced in case of Majorana neutrinos
  by a factor 2 compared to that in case of Dirac neutrinos.

\item The tensor NP contributions do not contribute to the
  differential decay rate for Majorana neutrinos, while for Dirac
  neutrinos such rate gets enhanced.  However, from
  Figs.\ref{fig:Pi2Pfnn-NP} it is clear that such enhancement is not
  substantial when compared with effects from other NP possibilities.
  Therefore, tensor NP interaction would be harder to probe than the
  other NP possibilities.

\end{enumerate}

\subsection{Patterns in $\boldsymbol{s}$ distributions of $\boldsymbol{B \to K \, \nu \, \overline{\nu}}$ decay in presence of NP}

\begin{figure*}[hbtp]
\centering%
\includegraphics[width=0.8\linewidth,keepaspectratio]{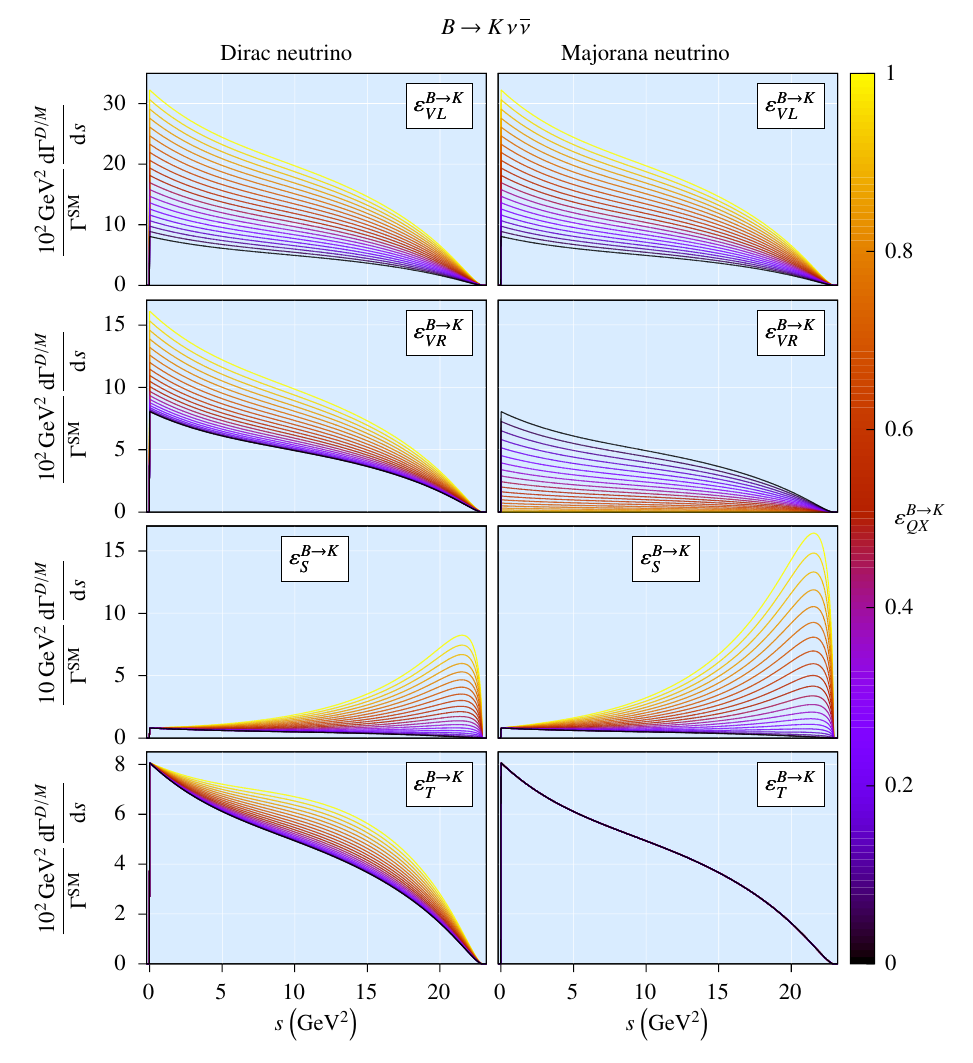}%
\caption{Comparison of various generic NP contributions in $B \to K \,
  \nu_\ell \, \overline{\nu}_\ell$ decay.  We use the scalar and
  tensor NP parameters $\varepsilon_{S}$ and $\varepsilon_{T}$ as
  defined in Eq.~\eqref{eq:ST-NP-parameters}. % Note that here
  $\Gamma_{\SM}$ denotes the partial decay rate of $B \to K \,
  \nu_\ell \, \overline{\nu}_\ell$ in the SM.  Also note the different
  overall normalisation factors used to show the distributions more
  clearly.} %
\label{fig:B2Knn-NP}
\end{figure*}

The discussion in section~\ref{subsec:generic-trends} is applicable to
all $\mathcal{P}_i \to \mathcal{P}_f \, \nu_\ell \,
\overline{\nu}_\ell$ decays but under the set of strong assumptions
which may not be satisfied in real processes. In particular, the form
factor $F_\textrm{SM}$ is known to be in general $s$-dependent, as
shown by the QCD lattice calculations. To study the effect of $s$
dependence of $F_\textrm{SM}$ on the distinguishing features of Dirac
and Majorana neutrinos, let us analyse as an example the decay $B \to
K \nu \overline{\nu}$, relaxing the assumptions we had made in
Sec.~\ref{subsec:generic-trends}. In particular, we use here the
functional form of $B \to K$ form factors in the SM as given in
Ref.~\cite{Bouchard:2013eph} and consider the full form of
Eqs.~\eqref{eq:D-dist-NP} and \eqref{eq:M-dist-NP} (although we have
continued to assume the unknown NP form factors $\varepsilon_{X}^{B
  \to K}$ to be independent of $s$ for simplicity).

Without the simplifying assumptions of
section~\ref{subsec:generic-trends}, the detailed shape of $s$
distributions changes as shown in Fig.~\ref{fig:B2Knn-NP}, but they
remain visibly different for Dirac and Majorana neutrino
possibilities. We note the following salient features noticeable in
Fig.~\ref{fig:B2Knn-NP}.
\begin{enumerate}
\item The general trends noted before in context of
  Fig.~\ref{fig:Pi2Pfnn-NP} are also clearly seen in
  Fig.~\ref{fig:B2Knn-NP}.  The presence of NP contributions, except
  the left-handed vector NP contribution, do indeed lead to different
  and distinguishable distributions for Dirac and Majorana neutrinos.

\item The distributions in Fig.~\ref{fig:B2Knn-NP} are significantly
  altered comparing to the ones shown in Fig.~\ref{fig:Pi2Pfnn-NP} for
  larger values of $s$, in concordance with the enhancement of
  $F_\mathrm{\SM}$ with increasing $s$.

\item We note also that if $m_\nu$ is not neglected, the distributions
  plotted in Fig.~\ref{fig:B2Knn-NP} actually drop to zero at the
  minimal value of $s = 4\,m_\nu^2$ (when the factor $b$ in the
  numerator of Eq.~\eqref{eq:gamsm} vanishes). However, as $m_\nu$ is
  extremely tiny in comparison with other masses, such drop of
  distributions is very steep and so close to $s=0$ that it is
  unlikely to be observed experimentally.
\end{enumerate} 

We would like to also emphasise the fact that in
Figs.~\ref{fig:Pi2Pfnn-NP} and \ref{fig:B2Knn-NP}, the NP parameters
are shown to be positive just for the purpose of illustration. In
Eqs.~\eqref{eq:DM-dist-NP} and \eqref{eq:XD-NP}, it is clear that the
scalar and tensor NP contributions are insensitive to the sign of the
NP parameter. For left-handed vector NP contribution alone, both Dirac
and Majorana cases yield identical $s$ distributions, where as for
right-handed vector NP contribution alone, the Dirac and Majorana
cases would still have opposite behaviour. Therefore, our conclusions
regarding distinguishability of Dirac and Majorana nature from
observed $s$ distribution still holds true, in general, even for
negative NP parameters.

Study of the energy spectra of the decays with neutrino antineutrino
pair in the final state is very challenging as such decays have very
low branching ratios. Notably also the example considered in this
section, $B \to K \nu \overline{\nu}$, has yet to be observed,
although current experimental bound for it is already less than order
of magnitude worse than the prediction of its SM decay
rate~\cite{Blake:2016olu, Belle-II:2021rof}.

In order to estimate the accuracy of spectra measurement required to
see the difference between the Dirac and Majorana neutrinos, let us
define the quantity %
\begin{equation}
\Delta X = \frac{\modulus{X^D-X^M}}{X^D}.
\end{equation}
One can assume two scenarios. First, one can estimate that the maximal
size of neutrino NP couplings to be of the order of limit given in
Eq.~\eqref{eq:nplimkll}, so ${\cal O}(0.1)$.  The second scenario
assume that there is a fine tuning between NP couplings and they may
be large, ${\cal O}(1)$, as shown in
Fig.~\ref{fig:epsilon-VLVR-constraint}. This is possible only for
$\varepsilon_{VL}$ and $\varepsilon_{VR}$ as scalar and tensor
contributions to total decay branching ratios are always positive and
cannot cancel out with other terms.  It turns out that the relative
modification of spectrum shape, given by $\Delta X$, is of the similar
order as the size of the (dimensionless) coupling $\varepsilon_X$
themselves. This can be e.g. seen assuming that only
$\varepsilon_{VR}$ does not vanish. Then one has for every $\tilde s$
the fixed ratio
\begin{equation}
\Delta X = \frac{2\modulus{\varepsilon_{VR}}}{1+ \varepsilon_{VR}^2}.
\end{equation}
Given the rarity of considered decays and statistics necessary to
obtain its energy spectrum, the first scenario $\varepsilon_{VR} \sim
\Delta X\sim {\cal O}(0.1)$ is unlikely to be observed at least in the
near future. However, in an (optimistic!) scenario of fine tuned large
NP couplings, $\varepsilon_{VR} \sim \Delta X\sim {\cal O}(1)$, the
nature of neutrinos can become clear if only the decay spectrum is at
all measurable, even with small accuracy.  If it happens, it also
points out to the type of the neutrino NP interactions, as only the
right-handed vector $\varepsilon_{VR}$ couplings can simultaneously be
large and lead to difference between the Dirac and Majorana case.

One should also note that Figs.~\ref{fig:Pi2Pfnn-NP} and
\ref{fig:B2Knn-NP} were plotted by taking individual non-vanishing NP
neutrino couplings one at a time. In general, the real shape of
spectrum can depend on several parameters and differ from those
displayed in Figs.~\ref{fig:Pi2Pfnn-NP} and \ref{fig:B2Knn-NP}. In any
specific NP model the effective shape is calculable and if experiment
is able to measure it, fitting procedures can give the information on
model parameters. However, as usual just discovery of discrepancy with
the SM predictions does not immediately disclose what kind of NP
interactions may be involved and further studies with combining more
observables are necessary.

As the experimental accuracy of these $s$ distributions gets improved,
one could be optimistic that in future with enough of detected events,
such a study might become feasible.  We hope that the possibility of
discovering NP as well as distinguishing the Dirac and Majorana nature
of neutrinos based on their different NP behaviours, would lend
support to any proposal for further experimental exploration of such
rare decays as $B \to K \, \nu \, \overline{\nu}$ and $K \to \pi \,
\nu \, \overline{\nu}$.

\section{Conclusions}
\label{sec:conclusion}

In this paper, we have provided a general framework for analysing the
processes containing a neutrino antineutrino pair in the final state,
considering neutrinos to be either Dirac or Majorana fermions.  We
emphasise the fact that the quantum mechanically identical nature of
Majorana neutrino and antineutrino does not depend on the size of
their mass.  Thus, the different statistical properties of the Dirac
and Majorana neutrinos can lead, at least in the New Physics models
with non-standard neutrino couplings, to significantly different
predictions for the decay rates and for the spectra of visible
particles.  Such differences are (at least in principle) not
proportional to the tiny neutrino masses, unlike the huge suppression
inherently present in the neutrino-mediated Lepton Number Violating
processes.

We argue that within the SM due to the particular structure of the
neutrino vector-like left-handed only couplings, such not-suppressed
effects of different statistics unfortunately still disappear when
integrating over the phase space of final state neutrinos (excluding
eventually some special kinematic scenarios where the neutrino momenta
can be reconstructed or inferred, see Ref.~\cite{Kim:2021dyj} for more
details).  However, we point out that, when New Physics effects are
taken into consideration, the difference between Dirac and Majorana
neutrino possibilities could survive even if one neglects neutrino
mass dependent terms and integrates over the phase space of final
state neutrinos.

We explicitly illustrate the general analysis by examples of neutral
current processes with the 2-body or 3-body final states.  Using our
example processes $Z \to \nu \, \overline{\nu}$ and $\mathcal{P}_i \to
\mathcal{P}_f \, \nu \, \overline{\nu}$ decays, we show how the
experimental bounds on the NP couplings differ between Dirac neutrino
and Majorana neutrino possibilities.  We also point out that if the NP
terms are substantial comparing to SM couplings and the differential
decay rate or equivalently the missing mass-square distribution for
$\mathcal{P}_i \to \mathcal{P}_f \, \nu \, \overline{\nu}$ decay can
be measured accurately enough, the shape of the spectrum can directly
help discern whether neutrinos are Dirac or Majorana fermions, at
least in NP models were there are no new light exotic states which
could also escape detection.

\section*{Acknowledgement}
%\begin{acknowledgement}
We would like to thank M.V.N.~Murthy for helpful discussions.  The
work of CSK is supported by NRF of Korea\\(NRF-2021R1A4A2001897 and
NRF-2022R1I1A1A01055643).  The work of JR is supported by the Polish
National Science Centre under the grant number
DEC-2016/23/G/ST2/04301.  The work of DS is supported by the Polish
National Science Centre under the grant number
DEC-2019/35/B/ST2/02008.  JR would also like to thank to CERN for the
hospitality during his stay there.
%\end{acknowledgement}


\begin{thebibliography}{99}

\bibitem{Fukuda:1998mi} Y.~Fukuda \textit{et al.} [Super-Kamiokande],
  ``Evidence for oscillation of atmospheric neutrinos, '' Phys.  Rev.
  Lett.  \textbf{81}, 1562-1567 (1998) doi:10.1103/PhysRevLett.81.1562
  [arXiv:hep-ex/9807003 [hep-ex]].

\bibitem{Ahmad:2002jz} Q.~R.~Ahmad \textit{et al.} [SNO], ``Direct
  evidence for neutrino flavor transformation from neutral current
  interactions in the Sudbury Neutrino Observatory, '' Phys.  Rev.
  Lett.  \textbf{89}, 011301 (2002) doi:10.1103/PhysRevLett.89.011301
  [arXiv:nucl-ex/0204008 [nucl-ex]].

\bibitem{Maki:1962mu} Z.~Maki, M.~Nakagawa and S.~Sakata, ``Remarks on
  the unified model of elementary particles, '' Prog.  Theor.  Phys.
  \textbf{28}, 870-880 (1962) doi:10.1143/PTP.28.870

\bibitem{Pontecorvo:1967fh} B.~Pontecorvo, ``Neutrino Experiments and
  the Problem of Conservation of Leptonic Charge, '' Zh.  Eksp.  Teor.
  Fiz.  \textbf{53}, 1717-1725 (1967) [Sov.\ Phys.\ JETP \textbf{26}
    984-988 (1968)]
  http://www.jetp.ras.ru/cgi-bin/e/index/e/26/5/p984?a=list

\bibitem{ParticleDataGroup:2020ssz} P.~A.~Zyla \textit{et al.}
  [Particle Data Group], ``Review of Particle Physics, '' PTEP
  \textbf{2020}, no.8, 083C01 (2020) and 2021 update.
  doi:10.1093/ptep/ptaa104

\bibitem{deGouvea:2016qpx} A.~de Gouv\^ea, ``Neutrino Mass Models, ''
  Ann.  Rev.  Nucl.  Part.  Sci.  \textbf{66}, 197-217 (2016)
  doi:10.1146/annurev-nucl-102115-044600

\bibitem{Cai:2017jrq} Y.~Cai, J.~Herrero-Garc\'\i{}a, M.~A.~Schmidt,
  A.~Vicente and R.~R.~Volkas, ``From the trees to the forest: a
  review of radiative neutrino mass models, '' Front.  in Phys.
  \textbf{5}, 63 (2017) doi:10.3389/fphy.2017.00063 [arXiv:1706.08524
    [hep-ph]].

\bibitem{Majorana:1937vz} E.~Majorana, ``Teoria simmetrica
  dell\textquoteright{}elettrone e del positrone, '' Nuovo Cim.
  \textbf{14}, 171-184 (1937) doi:10.1007/BF02961314

\bibitem{Racah:1937qq} G.~Racah, ``On the symmetry of particle and
  antiparticle, '' Nuovo Cim.  \textbf{14}, 322-328 (1937)
  doi:10.1007/BF02961321

\bibitem{Furry:1938zz} W.~H.~Furry, ``Note on the Theory of the
  Neutral Particle, '' Phys.  Rev.  \textbf{54}, 56-67 (1938)
  doi:10.1103/PhysRev.54.56

\bibitem{Avignone:2007fu} F.~T.~Avignone, III, S.~R.~Elliott and
  J.~Engel, ``Double Beta Decay, Majorana Neutrinos, and Neutrino
  Mass, '' Rev.  Mod.  Phys.  \textbf{80}, 481-516 (2008)
  doi:10.1103/RevModPhys.80.481 [arXiv:0708.1033 [nucl-ex]].


\bibitem{Blaum:2020ogl} K.~Blaum, S.~Eliseev, F.~A.~Danevich,
  V.~I.~Tretyak, S.~Kovalenko, M.~I.~Krivoruchenko, Y.~N.~Novikov and
  J.~Suhonen, ``Neutrinoless Double-Electron Capture, '' Rev.  Mod.
  Phys.  \textbf{92}, 045007 (2020) doi:10.1103/RevModPhys.92.045007
  [arXiv:2007.14908 [hep-ph]].

\bibitem{Lee:2021hnx} M.~Lee and M.~MacKenzie, ``Muon to Positron
  Conversion,'' Universe \textbf{8}, no.4, 227 (2022)
  doi:10.3390/universe8040227 [arXiv:2110.07093 [hep-ex]].

\bibitem{Grifols:1996fk} J.~A.~Grifols, E.~Masso and R.~Toldra,
  ``Majorana neutrinos and long range forces, '' Phys.  Lett.  B
  \textbf{389}, 563-565 (1996) doi:10.1016/S0370-2693(96)01304-4
         [arXiv:hep-ph/9606377 [hep-ph]].

\bibitem{Ferrer:1998ju} F.~Ferrer, J.~A.~Grifols and M.~Nowakowski,
  ``Long range forces induced by neutrinos at finite temperature, ''
  Phys.  Lett.  B \textbf{446}, 111-116 (1999)
  doi:10.1016/S0370-2693(98)01489-0 [arXiv:hep-ph/9806438 [hep-ph]].

\bibitem{Segarra:2020rah} A.~Segarra and J.~Bernab\'eu, ``Absolute
  neutrino mass and the Dirac/Majorana distinction from the weak
  interaction of aggregate matter, '' Phys.  Rev.  D \textbf{101},
  no.9, 093004 (2020) doi:10.1103/PhysRevD.101.093004
  [arXiv:2001.05900 [hep-ph]].

\bibitem{Costantino:2020bei} A.~Costantino and S.~Fichet, ``The
  Neutrino Casimir Force, '' JHEP \textbf{09}, 122 (2020)
  doi:10.1007/JHEP09(2020)122 [arXiv:2003.11032 [hep-ph]].

\bibitem{Ghosh:2019dmi} M.~Ghosh, Y.~Grossman and W.~Tangarife,
  ``Probing the two-neutrino exchange force using atomic parity
  violation, '' Phys.  Rev.  D \textbf{101}, no.11, 116006 (2020)
  doi:10.1103/PhysRevD.101.116006 [arXiv:1912.09444 [hep-ph]].

\bibitem{Bolton:2020xsm} P.~D.~Bolton, F.~F.~Deppisch and C.~Hati,
  ``Probing new physics with long-range neutrino interactions: an
  effective field theory approach, '' JHEP \textbf{07}, 013 (2020)
  doi:10.1007/JHEP07(2020)013 [arXiv:2004.08328 [hep-ph]].

\bibitem{Xu:2021daf} X.~j.~Xu and B.~Yu, ``On the short-range behavior
  of neutrino forces beyond the Standard Model: from 1/r$^{5}$ to
  1/r$^{4}$, 1/r$^{2}$, and 1/r, '' JHEP \textbf{02}, 008 (2022)
  doi:10.1007/JHEP02(2022)008 [arXiv:2112.03060 [hep-ph]].
  
\bibitem{Ghosh:2022nzo} M.~Ghosh, Y.~Grossman, W.~Tangarife, X.~J.~Xu
and B.~Yu, ``Neutrino forces in neutrino backgrounds,''
[arXiv:2209.07082 [hep-ph]].

\bibitem{Cvetic:2010rw} G.~Cvetic, C.~Dib, S.~K.~Kang and C.~S.~Kim,
  ``Probing Majorana neutrinos in rare $K$ and $D, D_s, B, B_c$ meson
  decays,'' Phys.  Rev.  D \textbf{82}, 053010 (2010)
  doi:10.1103/PhysRevD.82.053010 [arXiv:1005.4282 [hep-ph]].

\bibitem{CUORE:2021mvw} D.~Q.~Adams \textit{et al.} [CUORE], ``Search
  for Majorana neutrinos exploiting millikelvin cryogenics with
  CUORE,'' Nature \textbf{604}, no.7904, 53-58 (2022)
  doi:10.1038/s41586-022-04497-4 [arXiv:2104.06906 [nucl-ex]].

\bibitem{Kayser:1981nw} B.~Kayser and R.~E.~Shrock, ``Distinguishing
  Between Dirac and Majorana Neutrinos in Neutral Current Reactions,
  '' Phys.  Lett.  B \textbf{112}, 137-142 (1982)
  doi:10.1016/0370-2693(82)90314-8

\bibitem{Rosen:1982pj} S.~P.~Rosen, ``Analog of the Michel Parameter
  for Neutrino - Electron Scattering: A Test for Majorana Neutrinos,
  '' Phys.  Rev.  Lett.  \textbf{48}, 842 (1982)
  doi:10.1103/PhysRevLett.48.842

\bibitem{Dass:1984qc} G.~V.~Dass, ``Distinguishing between Dirac and
Majorana neutrinos in neutrino-electron scattering,'' Phys. Rev. D
\textbf{32}, 1239 (1985) doi:10.1103/PhysRevD.32.1239

\bibitem{Rodejohann:2017vup} W.~Rodejohann, X.~J.~Xu and C.~E.~Yaguna,
``Distinguishing between Dirac and Majorana neutrinos in the presence
of general interactions,'' JHEP \textbf{05}, 024 (2017)
doi:10.1007/JHEP05(2017)024 [arXiv:1702.05721 [hep-ph]].

\bibitem{Millar:2018hkv} A.~Millar, G.~Raffelt, L.~Stodolsky and
E.~Vitagliano, ``Neutrino mass from bremsstrahlung endpoint in
coherent scattering on nuclei,'' Phys. Rev. D \textbf{98}, no.12,
123006 (2018) doi:10.1103/PhysRevD.98.123006 [arXiv:1810.06584
[hep-ph]].

\bibitem{Kayser:1982br} B.~Kayser, ``Majorana Neutrinos and their
  Electromagnetic Properties, '' Phys.  Rev.  D \textbf{26}, 1662
  (1982) doi:10.1103/PhysRevD.26.1662

\bibitem{Schechter:1981hw} J.~Schechter and J.~W.~F.~Valle, ``Majorana
Neutrinos and Magnetic Fields,'' Phys. Rev. D \textbf{24}, 1883-1889
(1981) [erratum: Phys. Rev. D \textbf{25}, 283 (1982)]
doi:10.1103/PhysRevD.25.283

\bibitem{Shrock:1982jh} R.~E.~Shrock, ``Implications of Neutrino
  Masses and Mixing for the Mass, Width, and Decays of the $Z$,''
  eConf \textbf{C8206282}, 261-263 (1982) ITP-SB-82-52.

\bibitem{Ma:1989jpa} E.~Ma and J.~T.~Pantaleone,
  ``Heavy-Majorana-neutrino production,'' Phys.  Rev.  D \textbf{40},
  2172-2176 (1989) doi:10.1103/PhysRevD.40.2172

\bibitem{Nieves:1985ir} J.~F.~Nieves and P.~B.~Pal, ``Majorana
  neutrinos and photinos from kaon decay,'' Phys.  Rev.  D
  \textbf{32}, 1849-1852 (1985) doi:10.1103/PhysRevD.32.1849

\bibitem{Chhabra:1992be} T.~Chhabra and P.~R.~Babu, ``Way to
  distinguish between Majorana and Dirac massive neutrinos in neutrino
  counting reactions,'' Phys.  Rev.  D \textbf{46}, 903-909 (1992)
  doi:10.1103/PhysRevD.46.903

\bibitem{Berryman:2018qxn} J.~M.~Berryman, A.~de Gouv\^ea, K.~J.~Kelly
  and M.~Schmitt, ``Shining light on the mass scale and nature of
  neutrinos with $e\gamma \to e\nu\overline{\nu}$,'' Phys.  Rev.  D
  \textbf{98}, no.1, 016009 (2018) doi:10.1103/PhysRevD.98.016009
         [arXiv:1805.10294 [hep-ph]].

\bibitem{Yoshimura:2013wva} M.~Yoshimura and N.~Sasao, ``Radiative
  emission of neutrino pair from nucleus and inner core electrons in
  heavy atoms,'' Phys.  Rev.  D \textbf{89}, no.5, 053013 (2014)
  doi:10.1103/PhysRevD.89.053013 [arXiv:1310.6472 [hep-ph]].

\bibitem{Kim:2021dyj} C.~S.~Kim, M.~V.~N.~Murthy and D.~Sahoo,
  ``Inferring the nature of active neutrinos: Dirac or Majorana?,''
  Phys.  Rev.  D \textbf{105}, no.11, 113006 (2022)
  doi:10.1103/PhysRevD.105.113006 [arXiv:2106.11785 [hep-ph]].

\bibitem{Goudzovski:2022vbt} E.~Goudzovski, D.~Redigolo, K.~Tobioka,
J.~Zupan, G.~Alonso-\'Alvarez, D.~S.~M.~Alves, S.~Bansal, M.~Bauer,
J.~Brod and V.~Chobanova, \textit{et al.} ``New Physics Searches at
Kaon and Hyperon Factories,'' doi:10.1088/1361-6633/ac9cee
[arXiv:2201.07805 [hep-ph]].

\bibitem{Denner:1992vza} A.~Denner, H.~Eck, O.~Hahn and J.~Kublbeck,
  ``Feynman rules for fermion number violating interactions,'' Nucl.
  Phys.  B \textbf{387}, 467-481 (1992)
  doi:10.1016/0550-3213(92)90169-C

\bibitem{Marquez:2022bpg} J.~M.~M\'arquez, G.~L.~Castro and P.~Roig,
  ``Michel parameters in the presence of massive Dirac and Majorana
  neutrinos,'' [arXiv:2208.01715 [hep-ph]].

\bibitem{Janot:2019oyi} P.~Janot and S.~Jadach, ``Improved Bhabha
  cross section at LEP and the number of light neutrino species,''
  Phys.  Lett.  B \textbf{803}, 135319 (2020)
  doi:10.1016/j.physletb.2020.135319 [arXiv:1912.02067 [hep-ph]].


\bibitem{NA62:2021zjw} E.~Cortina Gil \textit{et al.} [NA62],
  ``Measurement of the very rare $K^+ \to \pi^+ \nu \overline{\nu}$
  decay,'' JHEP \textbf{06}, 093 (2021) doi:10.1007/JHEP06(2021)093
  [arXiv:2103.15389 [hep-ex]].
%

\bibitem{Buras:2015qea} A.~J.~Buras, D.~Buttazzo, J.~Girrbach-Noe and
  R.~Knegjens, ``$ {K}^{+}\to {\pi}^{+}\nu \overline{\nu} $ and $
  {K}_L\to {\pi}^0\nu \overline{\nu} $ in the Standard Model: status
  and perspectives,'' JHEP \textbf{11}, 033 (2015)
  doi:10.1007/JHEP11(2015)033 [arXiv:1503.02693 [hep-ph]].
%

\bibitem{Blake:2016olu} T.~Blake, G.~Lanfranchi and D.~M.~Straub,
  ``Rare $B$ Decays as Tests of the Standard Model,'' Prog.  Part.
  Nucl.  Phys.  \textbf{92}, 50-91 (2017)
  doi:10.1016/j.ppnp.2016.10.001 [arXiv:1606.00916 [hep-ph]].
%

\bibitem{Belle-II:2021rof} F.~Abudin\'en \textit{et al.} [Belle-II],
  ``Search for $B^+ \to K^+ \, \nu \, \overline{\nu}$ Decays Using an
  Inclusive Tagging Method at Belle II,'' Phys.  Rev.  Lett.
  \textbf{127}, no.18, 181802 (2021)
  doi:10.1103/PhysRevLett.127.181802 [arXiv:2104.12624 [hep-ex]].
%
  
\bibitem{Kim:2018hlp} C.~S.~Kim, S.~C.~Park and D.~Sahoo, ``Angular
distribution as an effective probe of new physics in semihadronic
three-body meson decays,'' Phys. Rev. D \textbf{100}, no.1, 015005
(2019) doi:10.1103/PhysRevD.100.015005 [arXiv:1811.08190 [hep-ph]].
%


\bibitem{Bouchard:2013eph} C.~Bouchard \textit{et al.} [HPQCD], ``Rare
  decay $B \to K \ell^+ \ell^-$ form factors from lattice QCD,'' Phys.
  Rev.  D \textbf{88}, no.5, 054509 (2013) [erratum: Phys.  Rev.  D
    \textbf{88}, no.7, 079901 (2013)] doi:10.1103/PhysRevD.88.054509
  [arXiv:1306.2384 [hep-lat]].

\end{thebibliography}
\end{document}